\documentclass[aps,prd,twocolumn,showpacs,superscriptaddress,preprintnumbers,floatfix,nofootinbib]{revtex4-2}

\usepackage{amsmath}
\usepackage{hyperref,url}
\usepackage{graphicx,subfigure}
\usepackage[normalem]{ulem}
\usepackage{epsfig}
\usepackage{color}
\usepackage{multirow}
\usepackage{array}
\usepackage[utf8]{inputenc}
\usepackage[T1]{fontenc}
\usepackage[dvipsnames]{xcolor}
\usepackage{comment}
\usepackage{amssymb,amsfonts}
\usepackage{bm}
\usepackage{booktabs}
\usepackage{siunitx}

\hypersetup{colorlinks   = true,
            urlcolor     = blue,
            citecolor    = blue,
            linkcolor    = blue,
            menucolor    = blue,
            anchorcolor  = blue,
            filecolor    = blue}

\enlargethispage{\baselineskip}

\newcommand{\Msun}{M_\odot}
\newcommand{\AU}{\mathrm{AU}}

\newcommand{\vecr}{\mathbf r}
\newcommand{\vecR}{\mathbf R}

\newcommand{\hatw}{\hat{\mathbf w}}
\newcommand{\avg}[1]{\left\langle #1 \right\rangle}

\usepackage{tikz}

\definecolor{lime}{HTML}{A6CE39}
\DeclareRobustCommand{\orcidicon}{
	\begin{tikzpicture}
	\draw[lime, fill=lime] (0,0) 
	circle [radius=0.16] 
	node[white] {{\fontfamily{qag}\selectfont \tiny ID}};
	\draw[white, fill=white] (-0.0625,0.095) 
	circle [radius=0.007];
	\end{tikzpicture}
	\hspace{-2mm}
}
\foreach \x in {A, ..., Z}{\expandafter\xdef\csname orcid\x\endcsname{\noexpand\href{https://orcid.org/\csname orcidauthor\x\endcsname}
			{\noexpand\orcidicon}}
}

\begin{document}

\title{Gravity Probe-DM: The Gravitational Laboratory for Dark Matter}

\author{Yu-Dai Tsai\hspace{-1mm}\orcidA{}}
\email{y.tsai@sheffield.ac.uk, yudaitsai.academic@gmail.com}
\affiliation{School of Mathematical and Physical Sciences, University of Sheffield, Sheffield S3 7RH, UK}
\affiliation{The University of Manchester, Manchester M13 9PL, United Kingdom}
\affiliation{Los Alamos National Laboratory (LANL), Los Alamos, NM 87545, USA}

\author{Hayden R. Foote\hspace{-1mm}\orcidB{}}
\affiliation{Steward Observatory, University of Arizona, 933 North Cherry Avenue, Tucson, AZ 85721, USA}

\date{\today}

\begin{abstract}
Dark matter is inferred gravitationally across the Universe but has not been detected within the Solar System. The Sun inevitably focuses incident unbound dark matter into an irreducible downstream wake. Its structure encodes the incoming density and velocity distribution and, for wave dark matter, the de Broglie scale. We propose \emph{Gravity Probe--DM}, a heliocentric search using precision ranging between spacecrafts. The exact two-spacecraft observable is the differential wake acceleration; for a short baseline, it becomes the wake tidal tensor projected along the baseline. 

For a $2.5\times10^6\,\mathrm{km}$ baseline crossing a $0.1\,\mathrm{AU}$ coherent width at $30\,\mathrm{km\,s^{-1}}$, a one-percent excess gives a raw uniform-core range scale of $0.63\,\mathrm{pm}$ over $5.77\,\mathrm d$. While the projected tide remains coherent over an interval $T$, the free response grows as $T^2$, reaching $2.5\,\mathrm{nm}$  over one year. An illustrative low-lag dark-disk component carrying $20\%$ of the reference local density and reaching a $30\%$ focused contrast gives a one-year scale of about $15\,\mathrm{nm}$. These pre-fit response scales show that trajectory design can move the signal from sub-picometre to nanometre scales. 

A detection would provide a purely gravitational map of local dark matter and probe the flow that produced it, including particle versus wave focusing. The Sun supplies the lens, spacecraft sample the wake, and precision ranging reads out its gravitational imprint.
\end{abstract}

\maketitle
\tableofcontents

\section{Introduction}
\label{sec:introduction}

Dark matter remains one of the central unsolved problems in physics. The cosmic microwave background and the growth of large-scale structure provide evidence on cosmological scales \cite{Planck:2018vyg}. Gravitational lensing in galaxy clusters and colliding systems \cite{2013SSRv..177...75H}, galaxy rotation curves \cite{1979ARA&A..17..135F}, and stellar motions in the smallest dark-matter-dominated galaxies \cite{2019ARA&A..57..375S} provide independent evidence on smaller scales. Gravity has established that dark matter exists. It has not revealed what dark matter is.

Our goals are twofold: to detect dark matter gravitationally for the first time in the Solar System, far below galactic scales, and to infer its properties from the measured structure.

Identifying the underlying particle or field will likely require a direct-detection signal, or some other non-gravitational signature. The interpretation of such a signal, however, depends on the dark-matter distribution in the Solar System. The local density sets event rates. The velocity distribution shapes recoil spectra, annual modulation, and directional signatures in particle detectors; for ultralight fields, it also sets the linewidth and coherence time \cite{Green:2017dd,Freese:2013modulation,Foster:2018halo}. Solar gravitational focusing further alters the slow component of the laboratory-frame distribution \cite{Sikivie:2002bj,Lee:2013wza}. The dependence on the local profile is particularly direct for proposals to search for solar-bound or planetary-bound ultralight dark matter~\cite{Banerjee:2019xuy,Lasenby:2020goo,VanTilburg:2020jvl}, using space quantum sensors~\cite{Tsai:2021lly,Tsai:2025wuf}.
The Solar System is therefore not merely where direct-detection experiments operate; it is part of their astrophysical signal model.

Yet dark matter has never been mapped gravitationally on Solar-System scales. Existing searches have placed limits on smooth dark matter around the Sun, using planetary and asteroid dynamics to look for anomalous precession or excess enclosed mass \cite{Pitjev:2013sfa,Verma:2017ywb,Tsai:2021irw,Tsai:2022jnv}. These bounds are essential, but they mainly ask how much dark matter can be hidden in a smooth, nearly spherical distribution. They do not ask whether the Solar System contains a directional dark-matter structure.

Such a structure arises unavoidably from gravitational focusing. Dark matter falling through the solar potential is accelerated and deflected, producing a downstream enhancement along the incident flow. Solar focusing is therefore an \emph{irreducible target}: for any incident unbound component, the Sun perturbs the particle trajectories or wave phases through gravity alone. The amplitude and detailed profile remain model dependent. Wave interference can produce enhanced and depleted fringes, but the focusing-induced structure itself does not rely on nongravitational couplings.

The morphology of the focused perturbation depends on which part of the local phase-space distribution is being lensed. In the broad, phase-mixed halo, contributions from many incident velocities wash one another out and leave a shallow net contrast. A component concentrated in velocity space can produce a more directional response and, in some cases, a larger one. Streams, debris flows, and co-rotating populations are useful enhancement cases. Co-rotating dark components occur in simulations of low-inclination accretion and disk-galaxy formation \cite{Read:2008fh,2009MNRAS.397...44R}.
Their low solar-frame speed has also motivated dedicated direct-detection benchmarks \cite{Bruch:2008rx}. A subdominant dissipative component could form a thinner disk by a different mechanism \cite{Fan:2013yva}, although Gaia stellar kinematics already restricts important parts of that parameter space \cite{Schutz:2017tfp,Buch:2018qdr,2021A&A...653A..86W}. We use a dark disk as an optimistic, astrophysically contingent benchmark separate from the baseline signal.

If the incident component is wave-like, ray tracing is no longer sufficient: finite-wavelength propagation regularizes the particle caustic and may produce alternating enhanced and depleted regions on the de Broglie scale \cite{Kim:2021yyo}. In either description, the gravitational source relevant to this experiment is the solar-induced departure from the ambient population, not the full local dark-matter density. We write
\begin{equation}
    \rho_{\rm DM}(\mathbf r)
    =
    \rho_\infty+\delta\rho_w(\mathbf r),
    \qquad
    \delta\rho_w(\mathbf r)
    =
    \rho_{\rm comp}\,\delta_w(\mathbf r),
    \label{eq:intro_excess_density}
\end{equation}
where $\rho_{\rm comp}$ is the asymptotic density of the component being focused and $\delta_w$ is its fractional focusing contrast. Throughout, $\delta\rho_w$ denotes the focusing-induced excess, and $\delta\rho_{\rm pk}$, when used, denotes its peak positive value. 

Recent progress in asteroid astrometry, radar ranging and Solar-System ephemerides has revived the prospect of detecting dark matter close to home \cite{Tsai:2021irw,Tsai:2022jnv}. But a passive search is not the only option. Once a dark-matter flow model specifies a downstream direction and an approximate wake width, a spacecraft trajectory can be designed to cross or skim the corresponding region.

In this paper we propose such a search. We call the mission concept \emph{Gravity Probe--DM}: one or two precision-tracked spacecraft on heliocentric trajectories chosen to sample the solar-focused dark-matter wake~\cite{Sikivie:2002bj,Alenazi:2006wu,Lee:2013wza,Kim:2021yyo}. The signal is purely gravitational and requires no assumed non-gravitational coupling between dark matter and the Standard Model. Seto and Cooray studied compact primordial-black-hole fly-bys in space-based interferometers in both the direct and tidal regimes \cite{Seto:2004zu}. Here the target is an extended, directional density perturbation, and the local projected tidal tensor is the more useful observable.

A single spacecraft would search for anomalous acceleration, secular
orbital perturbations, or a localized fly-through residual. A
two-spacecraft formation offers a more local measurement: the
differential acceleration across a controlled baseline, and hence the
projection of the wake tidal tensor along that baseline. In this sense,
the formation acts as a Solar-System dark-matter gradiometer. A uniform
acceleration cancels at leading order, but the solar and planetary tidal
fields do not. They are many orders of magnitude larger than the
expected wake signal. The orbit model must account for planetary ephemerides and asteroid
perturbations. It must also include solar radiation pressure and thermal
recoil. Manoeuvres and the baseline history enter the fit, together with clock
and ranging noise. Asteroid encounters are a particular concern because
a poorly known mass or orbit can produce a localized gravitational
perturbation. Candidate encounters may require improved orbit and mass
information, optical follow-up, or avoidance in the trajectory design.
Because the science arc will likely lie well outside the ecliptic,
planetary and asteroid backgrounds may be smaller than for an ecliptic
trajectory, especially when close encounters are avoided. Their long-range
tidal fields must still be modelled.

The sensitivity is set as much by the trajectory as by the ranging
instrument. The spacecraft must reach the predicted wake direction and
remain within the region of appreciable contrast. It must also preserve a
useful baseline orientation while keeping non-gravitational acceleration
noise sufficiently low. A transverse passage through a narrow wake may last only days to weeks. For the reference transverse coherence width
$D_{\rm coh}=0.1\,\mathrm{AU}$ crossed at
$v_\perp=30\,\mathrm{km\,s^{-1}}$, the crossing duration is
$5.77\,\mathrm d$. A $19\,\mathrm d$ crossing requires either a smaller
transverse speed or a larger transverse coherence width, as quantified in
Sec.~\ref{sec:reach}.

The raw free-response displacement scales as $T^2$ while the projected tidal field remains coherent. An approximately axial or wake-skimming orbit
can keep the guiding centre close to the wake axis for much longer than a transverse fly-through. A broad low-speed component can also relax the required transverse speed. In the idealized one-year coherent limit, the one-percent uniform-core normalization rises from $0.63\,\mathrm{pm}$ to $2.5\,\mathrm{nm}$. The low-lag dark-disk benchmark developed in Sec.~\ref{sec:long_dwell_dark_disk} reaches a raw scale of about $15\,\mathrm{nm}$ for the same baseline. These numbers show that the short-crossing benchmark is a conservative starting point for the concept.

The dark-disk case changes the target geometry more than it changes
the spacecraft dynamics. Its solar-frame lag vector fixes a wake axis
that need not coincide with the reference halo axis. The broader focused
profile also changes the useful impact parameter and dwell time. The
orbital plane and encounter epoch should be recomputed for the chosen disk
model, although the same general families of transverse and wake-skimming
trajectories remain applicable. The smooth Galactic potential of the disk
also leaves a heliocentric tidal term. This term is too weak to drive
transfer or targeting design, but it belongs in the precision propagation
and covariance model alongside the solar-focused wake.

Even with a well-reconstructed trajectory, wake density and width remain
correlated with the encounter geometry. The impact parameter and baseline
projection affect the amplitude, while the transverse speed sets the
coherent duration. A realistic sensitivity estimate must include these
correlations.

The wake profile, rather than a single amplitude, is the main science
observable. In a particle template, the ridge direction tracks the bulk
velocity. Its transverse extent and contrast respond to the velocity
dispersion and component density. A wave template adds mass-dependent
diffraction and interference, so its spatial pattern need not resemble a
broadened particle caustic. With sufficient sampling, a gravitational
measurement could constrain both the location of dark matter in the Solar
System and the phase-space component that produced the signal. Once the
mission sensitivity has been established with the same covariance model used
in the search, a null result would bound specified stream, dark-disk, and
wave-DM templates. The numerical calculations presented below test the
analytic mass and density scalings.

The remainder of the manuscript is arranged as follows. Section~\ref{sec:wake}
defines the wake geometry and summarizes particle and wave focusing.
Section~\ref{sec:mission} introduces the mission architecture and observables;
Sec.~\ref{sec:grace} reviews the relevant heritage from GRACE, GRACE Follow-On,
and LISA Pathfinder. Sections~\ref{sec:signals} and~\ref{sec:ranging} develop
the analytic signal estimates and differential-ranging response.
Section~\ref{sec:reach} addresses trajectory, baseline, and dwell-time
optimization. Section~\ref{sec:long_dwell_dark_disk} then develops the
long-dwell enhancement and the low-lag dark-disk benchmark, including the
smooth disk tide in the propagated trajectory. Section~\ref{sec:numerical_validation}
presents numerical checks of the leading scalings, and
Sec.~\ref{sec:backgrounds} describes the principal backgrounds and
discriminants. Section~\ref{sec:conclusion} summarizes the results and the
requirements for a realistic mission-sensitivity study. Its final two
subsections extend the concept to a network of differential links
(Sec.~\ref{sec:probe_network}) and explain how the recovered morphology can
constrain the density, kinematics, and wave scale of the incident component
(Sec.~\ref{sec:dm_property_implications}).

\section{Irreducible target: solar-focused dark matter}
\label{sec:wake}

We call solar focusing an irreducible target because its existence is
guaranteed by the Sun's gravity for every incident unbound dark-matter
component. This statement does not guarantee detectability, nor does it
require the density perturbation to be positive at every point: coherent
wave propagation can generate depleted fringes. It does guarantee a
calculable departure from the unfocused distribution. Gravity Probe--DM
is designed to measure that departure.

We consider an unbound dark-matter component with asymptotic density
$\rho_{\rm comp}$ and solar-frame velocity distribution
$f_\infty(\mathbf u)$. Its bulk velocity defines the downstream wake axis,
$\mathbf v_\infty=v_\infty\hat{\mathbf w}$. In heliocentric ecliptic
coordinates,
\begin{equation}
    \hat{\mathbf w}
    =
    \left(
        \cos\theta_w\cos\phi_w,
        \cos\theta_w\sin\phi_w,
        \sin\theta_w
    \right),
    \label{eq:wake_axis}
\end{equation}
where $\theta_w$ and $\phi_w$ are the downstream ecliptic latitude and
longitude. The apparent arrival direction of the flow is
$-\hat{\mathbf w}$. For any heliocentric position $\mathbf r$, we define
\begin{equation}
    s\equiv\mathbf r\cdot\hat{\mathbf w},
    \qquad
    \mathbf r_\perp\equiv\mathbf r-s\hat{\mathbf w},
    \qquad
    r_\perp\equiv|\mathbf r_\perp|.
    \label{eq:wake_coordinates}
\end{equation}

The gravitational source of interest is the focusing-induced departure from
the unfocused density~
\cite{Sikivie:2002bj,Alenazi:2006wu,Lee:2013wza,Kim:2021yyo},
\begin{equation}
\begin{aligned}
    \rho_{\rm DM}(\mathbf r)
    &=
    \rho_\infty+\delta\rho_w(\mathbf r),
    \\
    \delta\rho_w(\mathbf r)
    &=
    \rho_{\rm comp}\,\delta_w(\mathbf r)
    \\
    &=
    \delta\rho_{\rm pk}\,f_w(\mathbf r).
\end{aligned}
\label{eq:wave_density_template}
\end{equation}
Here $\delta_w$ is the fractional focusing contrast of the selected
component, $\delta\rho_{\rm pk}$ is its peak positive excess, and $f_w$ is a
dimensionless profile normalized to unit positive maximum. The profile may
contain depleted regions for wave dark matter. The full ambient density
$\rho_\infty$ is not treated as a finite coherent overdensity.

For particle dark matter, $f_w$ is obtained by transporting the asymptotic
phase-space distribution through the solar potential. For wave dark matter,
it is obtained from coherent propagation and averaging over the incident
velocity distribution. Each benchmark fixes the density normalization and
wake direction within one consistent model. The same model also fixes the
transverse width and longitudinal structure. We denote by $w_{\rm phys}(s)$ a
stated width extracted from the calculated profile; navigation and pointing
uncertainties are kept separate.

The wake potential and tidal tensor satisfy
\begin{equation}
    \nabla^2\Phi_w
    =
    4\pi G\,\delta\rho_w,
    \qquad
    \mathcal E^w_{ij}
    \equiv
    \partial_i\partial_j\Phi_w.
    \label{eq:wave_tidal_tensor}
\end{equation}
The ranging observable depends on the projection of this full finite-profile
Hessian along the spacecraft baseline (not on the density alone). The particle
and wave constructions, approximate width scalings, and benchmark conventions
are given in Appendix~\ref{app:wake_focusing}.

\section{Mission concept}
\label{sec:mission}

\begin{figure}[h]
    \centering
    \includegraphics[width=\columnwidth]{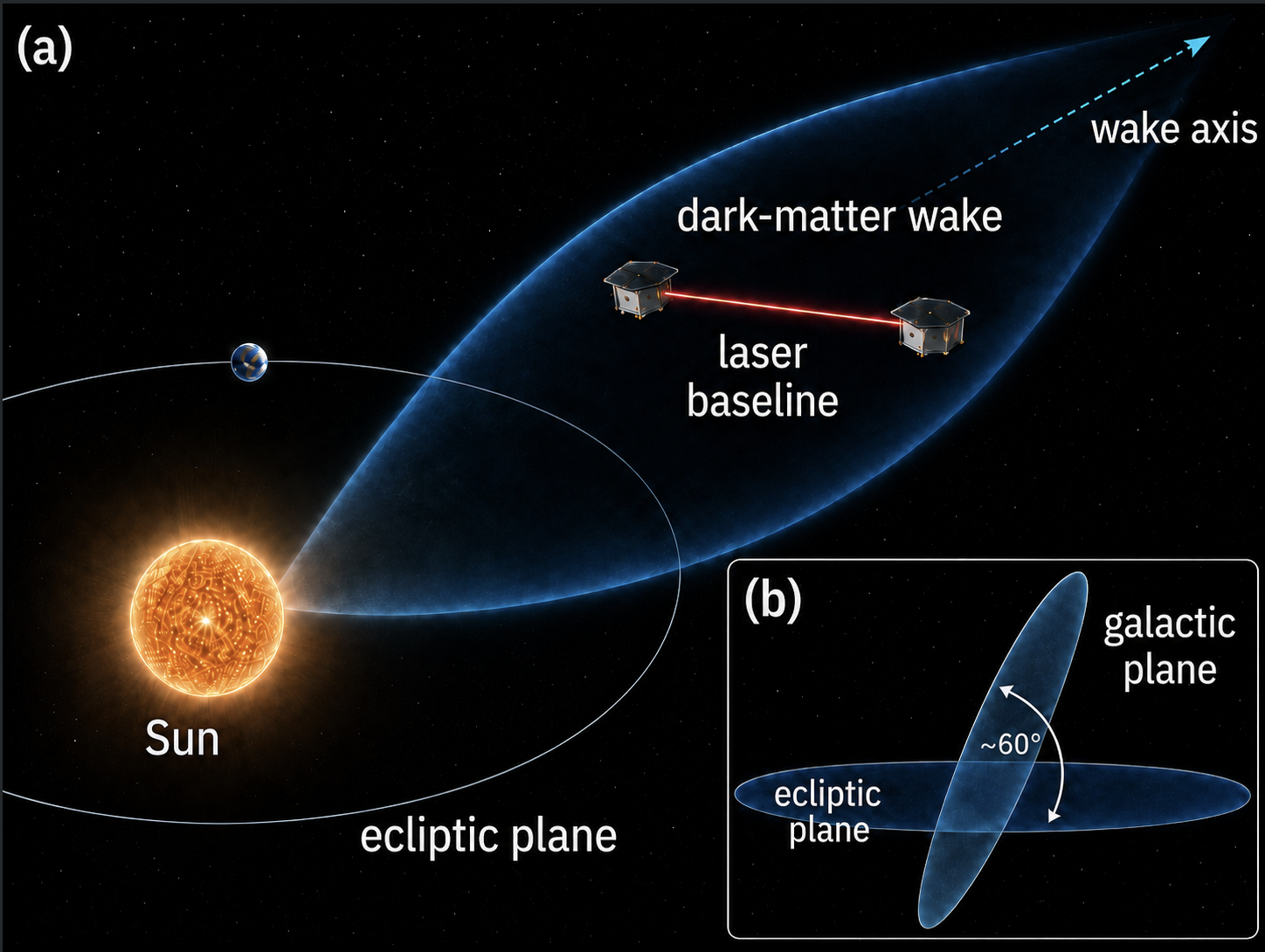}
\caption{\textbf{Gravity Probe--DM mission concept.}
\textbf{(a)} Schematic of a heliocentric two-spacecraft experiment designed to search for the solar dark-matter wake. Galactic dark matter is gravitationally focused by the Sun, producing a downstream overdensity, shown in blue, whose orientation defines the wake axis. Two spacecraft traversing or skimming the wake are connected by a laser baseline, enabling a differential measurement of the wake-induced tidal field. The ecliptic plane and Earth's orbit are shown for orientation.
\textbf{(b)} Geometry of the ecliptic plane relative to the Galactic
plane. The two planes are inclined by approximately $60^\circ$, but
their mutual inclination does not directly determine the wake direction. The
actual wake axis must be obtained by transforming the specified
solar-frame flow velocity into the adopted ecliptic coordinate system. The illustration is schematic and not to scale.}
    \label{fig:illustration}
\end{figure}

Gravity Probe--DM is a solar mission designed to cross the downstream wake of a selected DM flow component. The minimal concept uses one precision-tracked spacecraft. A pair separated by a designed baseline provides the stronger differential measurement.

The proposed analysis is:
\begin{enumerate}
    \item specify a self-consistent dark-matter phase-space component, whose solar-frame bulk velocity fixes the downstream wake direction and whose density and velocity distribution determine the focused profile;

    \item for wave dark matter, specify the mass controlling the diffraction and interference scales;

    \item compute the finite focusing-induced excess density and its gravitational field;

    \item propagate a feasible heliocentric spacecraft formation through the resulting wake template;

    \item construct range, range-rate, Doppler, and differential-acceleration observables; 

    \item fit the wake simultaneously with Solar-System, spacecraft-force, and instrument parameters.
\end{enumerate}

For a single spacecraft, the wake contribution to the gravitational
acceleration is
\begin{equation}
    \mathbf A_w(\mathbf r,t)
    =
    -\nabla\Phi_w(\mathbf r,t),
    \label{eq:wake_acceleration}
\end{equation}
where $\Phi_w$ is sourced by the focusing-induced excess density
$\delta\rho_w$. In a Sun-centred orbit fit, the observable is the
spacecraft motion relative to the solar reference trajectory.

For two spacecraft, define
\begin{equation}
    \mathbf L(t)
    \equiv
    \mathbf r_2(t)-\mathbf r_1(t),
    ~~~
    L(t)=|\mathbf L(t)|,
    ~~~
    \hat{\mathbf L}(t)
    =
    \frac{\mathbf L(t)}{L(t)},
    \label{eq:baseline_definition}
\end{equation}
and
\begin{equation}
    \mathbf r_g(t)
    \equiv
    \frac{\mathbf r_1(t)+\mathbf r_2(t)}{2}.
    \label{eq:guiding_centre_definition}
\end{equation}
The exact wake-induced differential acceleration is
\begin{equation}
    \Delta a_i^w(t)
    =
    A_{w,i}(\mathbf r_2,t)
    -
    A_{w,i}(\mathbf r_1,t).
    \label{eq:exact_wake_difference}
\end{equation}
When the baseline is short compared with the local variation scale of
the wake field, a midpoint expansion gives
\begin{equation}
    \Delta a_i^w(t)
    \simeq
    -\mathcal E^w_{ij}(\mathbf r_g,t)L_j(t),
    \qquad
    \mathcal E^w_{ij}
    =
    \partial_i\partial_j\Phi_w .
    \label{eq:diff_accel}
\end{equation}
The corresponding line-of-sight contribution is
\begin{equation}
    \Delta a_\parallel^w(t)
    \simeq
    -L(t)\,
    \hat L_i(t)
    \mathcal E^w_{ij}(\mathbf r_g,t)
    \hat L_j(t).
    \label{eq:projected_tidal_signal}
\end{equation}
A single baseline measures one scalar projection of the wake Hessian.

A LISA-like separation,
$L=2.5\times10^6\,\mathrm{km}=0.0167\,\mathrm{AU}$,
is used as a reference scale.
The finite wake profile and formation dynamics set the main physical
trade-offs. Link performance and nuisance covariance must also enter the
final baseline choice.

\subsection{Connecting to existing missions: GRACE, GRACE Follow-On and LISA Pathfinder}
\label{sec:grace}

Two flight demonstrations provide the closest heritage for Gravity Probe--DM.
GRACE and its successor, GRACE Follow-On, showed that two spacecraft in
nearly the same low-Earth orbit can operate as a low--low
satellite-to-satellite gravity gradiometer. Changes in inter-spacecraft range
and range rate encode variations in Earth's gravity field (i.e., measuring Earth's multipole field) \cite{2004GeoRL..31.9607T}. GRACE Follow-On retained the microwave ranging
system and added the Laser Ranging Interferometer. It demonstrated laser
interferometric ranging over a separation of order $220\,{\rm km}$
\cite{Sheard:2012gsj,Abich:2019cci}.

LISA Pathfinder provides complementary heritage. Its two shielded test masses
were separated by $38\,{\rm cm}$ within a single spacecraft. The mission
demonstrated drag-free control and precision optical metrology, and it built
an in-flight model of residual forces on nominally free-falling reference
bodies \cite{2012CQGra..29l4014A,Armano:2016bkm,Armano:2018kix}. GRACE thus
supplies the ranging geometry and geodetic data-analysis analogy. LISA
Pathfinder supplies the inertial-reference and disturbance-reduction analogy.

Gravity Probe--DM would combine these lessons in a heliocentric setting. Unlike GRACE, it targets the much weaker tidal field of a predicted solar dark-matter wake. In the local short-baseline approximation, the wake-induced
line-of-sight differential acceleration is
\begin{equation}
    \Delta a_\parallel^w
    \simeq
    -L\,
    \hat L_i
    \mathcal E^w_{ij}
    \hat L_j ,
    \label{eq:grace_like_tidal}
\end{equation}
where $\mathcal E^w_{ij}$ is the wake Hessian and $\hat{\mathbf L}$ is
the baseline direction. The local expansion requires
$L\ll\ell_{\mathcal E}$, with the Hessian-variation scale defined in
Eq.~(\ref{eq:hessian_variation_scale}). If this condition is not
satisfied, the exact finite difference in
Eq.~(\ref{eq:exact_wake_difference}) must be used. For GRACE, the relevant
length scales were set by Earth-orbit altitude, ground-track sampling, and
terrestrial mass redistribution. For a solar dark-matter wake, the wake width
and spacecraft ephemeris set the corresponding scales. Wave dark matter adds
smoothing over $\lambda_{\rm dB}$. This permits baselines far larger than
GRACE's. The link must still meet acquisition and clock-transfer requirements,
and thermal recoil, radiation pressure, and ephemeris errors must remain below
the wake-induced differential acceleration.

The data product can become GRACE-like after detailed modelling. Raw range, range rate, or optical phase would be fitted with the solar and planetary gravity model. The same fit would include nongravitational accelerations and thermal forces. Manoeuvres, clock noise, and tracking systematics would enter with them. A candidate wake signal would appear as a coherent, baseline-projected tidal waveform. Its phase, sign, and duration would be tied to the spacecraft path through the predicted wake tube.

The two precedents remain useful in different ways. GRACE shows that weak
gravity signals can be recovered from differential inter-spacecraft tracking
after background subtraction. LISA Pathfinder shows that force and metrology
disturbances on free-falling references can be calibrated and cross-checked in
flight. Gravity Probe--DM is a distinct heliocentric experiment. It adopts
GRACE's differential-ranging observable and LISA Pathfinder's treatment of
the spacecraft as a shielded, calibrated inertial sensor.

\section{Analytic signal estimates}
\label{sec:signals}

\subsection{Three gravitational signal classes}
\label{sec:signal_classes}

It is useful to distinguish three gravitational signal classes relevant
to Solar-System searches. First, a compact object such as a primordial
black hole produces a localized encounter. Depending on the closest
approach relative to the detector baseline, the observable is either a
direct acceleration pulse or a tidal pulse. Previous work developed the
direct and tidal response regimes for space-based interferometers,
emphasizing low-frequency proof-mass noise and confusion from asteroid
encounters~\cite{Seto:2004zu}; a complementary approach proposed
searching for asteroid-mass primordial black holes through their
characteristic perturbations of Solar-System orbits
\cite{Tran:2023jci}.

Second, a smooth or slowly varying Solar-System dark-matter distribution produces a broad, approximately central perturbation. It changes the enclosed mass and drives secular orbital precession. Over longer intervals,
the same perturbation shifts the trajectories of planets, asteroids, and spacecrafts. Such signals are most naturally constrained through a global
orbit or ephemeris fit
\cite{Pitjev:2013sfa,Verma:2017ywb,Tsai:2021irw,Tsai:2022jnv}.

The signal considered primarily in this paper belongs to a third class.
A solar-focused wake is an extended, anisotropic excess density tied to
an incident dark-matter flow. Its local signature is not generally
described by a single compact mass or by a spherically symmetric density.
For a spacecraft pair, the short-baseline observable is the
trajectory-dependent projection
\begin{equation}
    \Delta a_\parallel^w
    \simeq
    -L\,
    \hat L_i
    \mathcal E^w_{ij}
    \hat L_j
\end{equation}
of the Hessian generated by the finite excess-density profile. The wake
amplitude and width are correlated with the encounter geometry. Impact
parameter and baseline orientation set the projection, while transverse
velocity sets the dwell time.

Although the solar-focused wake is the primary target of
Gravity Probe--DM, the same heliocentric tracking architecture would retain
sensitivity to the first two signal classes. A compact-object encounter would
leave an acceleration transient in the individual tracking data and a tidal
transient in the inter-spacecraft link. Their correlated appearance would
strengthen identification. A smooth Solar-System dark-matter component would
produce a long-duration perturbation in the spacecraft trajectories and range
history, to be constrained through the global orbit solution. The three
searches differ in template and timescale, and each needs its own nuisance
model. A trajectory optimized for the solar-focused wake need not be optimal
for the other cases. Even so, the same mission data could support all three
analyses. Gravity Probe--DM could thus serve more broadly as a gravitational
observatory for compact, smooth, and solar-focused dark structures in the
Solar System.

\subsection{Compact-source limit}

As a first limiting case, we approximate the focused excess density by an
effective compact mass $M_{\rm od}$ located on the wake axis,
\begin{equation}
    \vecR = R\hatw ,
    \qquad R=|\vecR| ,
\end{equation}
where $\hatw$ is the downstream wake direction. In this limit the perturbing
potential is that of a point source,
\begin{equation}
    \Phi_1(\vecr)
    =
    -\frac{G M_{\rm od}}{|\vecr-\vecR|},
    \label{eq:point_potential}
\end{equation}
with $\vecr$ the heliocentric position of the spacecraft and
$r=|\vecr|$. For $r\ll R$, the multipole expansion about the Sun is
\begin{equation}
\Phi_1(\vecr)=
-\frac{G M_{\rm od}}{R}
\left[
1+\frac{\vecr\cdot\vecR}{R^2}
+\frac{3(\vecr\cdot\vecR)^2-r^2R^2}{2R^4}
+\cdots
\right].
\label{eq:multipole}
\end{equation}

The constant term has no force. The dipole term represents a common
acceleration of the Sun and spacecraft produced by the distant
perturber. Because heliocentric motion is measured relative to the
solar position, the relevant acceleration is the difference between the
acceleration at the spacecraft and that at the Sun,

\begin{equation}
    \mathbf A_{\rm rel}(\vecr)
    =
    -\nabla\Phi_1(\vecr)
    +
    \nabla\Phi_1(\mathbf 0).
\end{equation}
Thus the leading measurable contribution in the far-field limit is the
quadrupolar, or tidal, term. Equivalently, the observable tidal potential is
\begin{equation}
    \Phi_{\rm tid}(\vecr)
    =
    -\frac{G M_{\rm od}}{2R^5}
    \left[
    3(\vecr\cdot\vecR)^2-r^2R^2
    \right]
    +\cdots ,
    \label{eq:tidal_potential_compact}
\end{equation}
which produces the relative acceleration
\begin{equation}
    A_{{\rm rel},i}
    \simeq
    \frac{G M_{\rm od}}{R^3}
    \left(
    3\hat R_i\hat R_j-\delta_{ij}
    \right) r_j ,
    \qquad
    \hat{\mathbf R}\equiv \frac{\vecR}{R}.
    \label{eq:tidal_accel_compact}
\end{equation}

For a bound heliocentric orbit with semimajor axis $a$, eccentricity $e$,
and mean motion
\begin{equation}
    n=\sqrt{\frac{G M_\odot}{a^3}},
\end{equation}
the induced apsidal precession scales as
\begin{equation}
    \avg{\dot\omega}_{\rm tide}
    \sim
    n
    \frac{M_{\rm od}}{M_\odot}
    \left(\frac{a}{R}\right)^3
    \mathcal F(e,\mathcal O;\hatw),
    \label{eq:tide_scaling}
\end{equation}
where $\mathcal O$ denotes the orientation of the orbit, for example its
inclination, node and argument of perihelion, and $\mathcal F$ is an
order-unity geometric factor. The $R^{-3}$ dependence is the asymptotic
tidal limit and should be used when the compact source is well separated
from the sampled orbit, $R\gg a$; for $R/a$ of only a few, the full orbit
integration gives order-unity corrections.

For a close fly-through, the tidal expansion is no longer appropriate. If the
spacecraft passes a compact excess mass with relative speed $v_{\rm rel}$ and
minimum separation $b_{\rm fly}$, the transverse impulse is
\begin{equation}
    \Delta v
    \sim
    \frac{2G M_{\rm od}}{v_{\rm rel} b_{\rm fly}},
\end{equation}
and the corresponding deflection angle is
\begin{equation}
    \Delta\psi
    \sim
    \frac{2G M_{\rm od}}{v_{\rm rel}^2 b_{\rm fly}} .
    \label{eq:impulse}
\end{equation}
These estimates apply to the compact-source limit. For an extended wake, the
same observables must be computed from the integrated gravitational field of
the density profile. A point-mass approximation is then inadequate.

\subsection{Extended wake limit}

For a finite wake, a single effective mass is not the most useful local
description. The relevant quantity is the tidal field generated by the
density profile. In this subsection we establish its characteristic scale
and relate it to orbital precession; the projected two-spacecraft response
is developed in Sec.~\ref{sec:ranging}.

We describe the wake by the focusing-induced excess density
\begin{equation}
    \delta\rho_w(\mathbf r)
    =
    \rho_{\rm comp}\delta_w(\mathbf r)
    =
    \delta\rho_{\rm pk}f_w(\mathbf r),
    \label{eq:extended_excess_density}
\end{equation}
using the conventions of Eq.~(\ref{eq:wave_density_template}). The
excess mass contained in a sampled volume $V$ is
\begin{equation}
    M_{\rm od}(V)
    =
    \int_V
    \delta\rho_w(\mathbf r)\,
    \mathrm d^3x.
    \label{eq:excess_mass}
\end{equation}

For analytic and numerical validations, consider the finite Gaussian toy
profile
\begin{equation}
    \delta\rho_w^{\rm toy}(s,\mathbf r_\perp)
    =
    \delta\rho_{\rm pk}
    \exp\left[
        -\frac{r_\perp^2}{2w^2}
    \right]
    \exp\left[
        -\frac{(s-s_0)^2}{2\ell^2}
    \right],
    \label{eq:gaussian_wake}
\end{equation}
where
\begin{equation}
    s=\mathbf r\cdot\hat{\mathbf w},
    \qquad
    \mathbf r_\perp
    =
    \mathbf r-s\hat{\mathbf w},
    \qquad
    r_\perp=|\mathbf r_\perp|.
\end{equation}
Here $\delta\rho_{\rm pk}$ is the peak excess density, $w$ is the
transverse Gaussian scale, and $\ell$ is the longitudinal Gaussian
scale.

Its potential and Hessian satisfy
\begin{equation}
    \nabla^2\Phi_w
    =
    4\pi G\,\delta\rho_w^{\rm toy},
    \qquad
    \mathcal E^w_{ij}
    =
    \partial_i\partial_j\Phi_w.
    \label{eq:extended_wake_poisson}
\end{equation}
Poisson's equation fixes the trace,
\begin{equation}
    \mathcal E^w_{ii}
    =
    4\pi G\,\delta\rho_w^{\rm toy},
\end{equation}
but not the individual tensor components. Those depend on the position
within the wake and on the aspect ratio $\ell/w$. Near the wake core, a
characteristic component therefore scales as
\begin{equation}
    \mathcal E^w_{\rm char}
    \sim
    4\pi G\,\delta\rho_{\rm pk},
    \label{eq:extended_tidal_scale}
\end{equation}
up to a dimensionless factor determined by the complete density profile
and by the measurement projection.

The compact-source precession estimate in
Eq.~(\ref{eq:tide_scaling}) can be written in terms of the same tidal
scale, since
\begin{equation}
    n
    \frac{M_{\rm od}}{M_\odot}
    \left(\frac{a}{R}\right)^3
    =
    \frac{1}{n}
    \frac{GM_{\rm od}}{R^3}.
\end{equation}
More generally, a weak tidal field that varies slowly across the sampled
orbit produces an apsidal precession of order
\begin{equation}
    \avg{\dot\omega}_{\rm ext}
    \sim
    \frac{\mathcal E_{\rm orb}}{n}
    \mathcal F_{\rm ext},
    \label{eq:extended_precession_scaling}
\end{equation}
where $\mathcal E_{\rm orb}$ is a characteristic orbit-averaged tidal
amplitude. The factor $\mathcal F_{\rm ext}$ captures the eccentricity and
orbital orientation, together with the wake geometry. For an orbit that
samples the wake core,
\begin{equation}
    \avg{\dot\omega}_{\rm ext}
    \sim
    \frac{4\pi G\,\delta\rho_{\rm pk}}{n}
    \mathcal F_{\rm ext}.
    \label{eq:extended_precession_core}
\end{equation}
If the tidal field changes substantially during one orbit, a single
secular precession rate is not generally adequate; the full
time-dependent acceleration from Eq.~(\ref{eq:gaussian_wake}) must then
be included in the orbit integration.

For two spacecraft, the weaker locality condition is that the baseline be short compared with the local variation scale of the wake. Equation (\ref{eq:diff_accel}) then gives
\begin{equation}
    |\Delta\mathbf a|
    \sim
    \mathcal E^w_{\rm char}L
    \sim
    4\pi G\,\delta\rho_{\rm pk}L,
    \label{eq:extended_diff_accel_scaling}
\end{equation}
again up to the tensor projection and wake geometry. Thus both the precession and the differential-ranging signals are linear in the tidal tensor, while the latter remains useful even when the spacecraft samples only a localized portion of the wake.

At distances large compared with both $w$ and $\ell$, the finite wake is
seen through its total excess mass
$M_{\rm od}^{\rm tot}=M_{\rm od}(\mathbb R^3)$, and the compact-source
results of the preceding subsection are recovered.

\section{Differential ranging}
\label{sec:ranging}

The wake contribution should first be evaluated using the exact finite
difference in Eq.~(\ref{eq:exact_wake_difference}). The local expression
in Eq.~(\ref{eq:projected_tidal_signal}) is used only when the baseline
is short compared with the spatial variation scale of the wake Hessian.

For a free relative coordinate with zero injected displacement and
velocity at $t_0$, the raw line-of-sight range response is
\begin{equation}
    \Delta L_{\rm raw}(t)
    =
    \int_{t_0}^{t}
    (t-t')\,
    \Delta a_\parallel^w(t')\,
    \mathrm dt'.
    \label{eq:deltar_general}
\end{equation}
If the projected tidal component and the baseline geometry are
approximately constant over an interval $T$, this reduces to
\begin{equation}
    |\Delta L_{\rm raw}|
    \simeq
    \frac{1}{2}
    L
    \left|
        \hat L_i
        \mathcal E^w_{ij}(\mathbf r_g)
        \hat L_j
    \right|
    T^2.
    \label{eq:constant_projected_tide_range}
\end{equation}

For normalization, consider a uniform spherical excess of density
$\delta\rho_{\rm core}$. Inside the sphere,
\begin{equation}
    \mathcal E^{\rm core}_{ij}
    =
    \frac{4\pi G\,\delta\rho_{\rm core}}{3}
    \delta_{ij}.
    \label{eq:spherical_core_tidal}
\end{equation}
The corresponding differential acceleration is
\begin{equation}
    \Delta a_i^{\rm core}
    =
    -\frac{4\pi G\,\delta\rho_{\rm core}}{3}
    L_i,
    \label{eq:spherical_core_diff_accel}
\end{equation}
and the raw short-arc range normalization is
\begin{equation}
    |\Delta L_{\rm raw}^{\rm core}|
    \simeq
    \frac{2\pi G\,\delta\rho_{\rm core}}{3}
    LT^2.
    \label{eq:deltar_uniform}
\end{equation}

For the reference crossing,
\begin{equation}
\begin{aligned}
    |\Delta L_{\rm raw}^{\rm core}|
    \simeq{}&
    6.3\times10^{-11}\,\mathrm m
    \left(
        \frac{\delta\rho_{\rm core}}
        {0.4\,\mathrm{GeV\,cm^{-3}}}
    \right)
    \left(
        \frac{L}{0.0167\,\mathrm{AU}}
    \right)
    \\
    &\times
    \left(
        \frac{T}{5.77\,\mathrm d}
    \right)^2.
\end{aligned}
    \label{eq:wave_delta_r_number}
\end{equation}
A one-percent excess,
$\delta\rho_{\rm core}
=4.0\times10^{-3}\,\mathrm{GeV\,cm^{-3}}$,
therefore gives
\begin{equation}
    |\Delta L_{\rm raw}^{\rm core}|
    \simeq
    0.63\,\mathrm{pm}
    \left(
        \frac{L}{0.0167\,\mathrm{AU}}
    \right)
    \left(
        \frac{T}{5.77\,\mathrm d}
    \right)^2.
    \label{eq:percent_excess_range_number}
\end{equation}

These expressions are raw normalizations. A physical wake is generally
neither spherical nor isotropic. Its projected Hessian can change in both
magnitude and sign along the trajectory. The exact acceleration difference
must be propagated through the formation dynamics and the state-estimation
model used for the data analysis.

If $\sigma_{\Delta L}$ denotes the post-fit uncertainty in the amplitude
of this same uniform-core template, an equivalent excess-density scale
may be defined by
\begin{equation}
    \delta\rho_{{\rm core},{\rm eq}}
    \sim
    \frac{3\sigma_{\Delta L}}
    {2\pi G L T^2}.
    \label{eq:rho_min}
\end{equation}
Equation~(\ref{eq:rho_min}) is an equivalent uniform-core
normalization. For a physical wake, the
mapping between a fitted range amplitude and $\delta\rho_{\rm pk}$ depends on
the finite-profile Hessian and the trajectory. Fitted spacecraft states and
nuisance covariance also affect that mapping.

\section{Mission reach and optimization}
\label{sec:reach}

The analytic estimates imply three optimization principles.

First, the trajectory must intersect the physical wake. The nominal
miss distance should be compared with the physical width
$w_{\rm phys}$ obtained from the calculated density profile, while
navigation and wake-direction uncertainties enter separately through
the targeting uncertainty $\sigma_b$. A robust targeting condition is
given in Eq.~(\ref{eq:trajectory_requirements}).

Second, the mission should maximize coherent dwell time. 
The relevant speed is the component of the guiding-centre velocity
transverse to the wake axis,
\begin{equation}
    v_\perp(t)
    \equiv
    \left|
    \dot{\mathbf r}_g
    -
    \bigl(\dot{\mathbf r}_g\!\cdot\!\hat{\mathbf w}\bigr)
    \hat{\mathbf w}
    \right|.
    \label{eq:transverse_velocity}
\end{equation}
To distinguish a Gaussian scale radius from a full diameter or a fitted
coherence interval, we denote by $D_{\rm coh}$ the effective
transverse distance over which the projected wake tide keeps comparable
amplitude and sign during a predominantly transverse passage. The
corresponding crossing time is
\begin{equation}
\begin{aligned}
    T_{\rm cross}
    &\simeq
    \frac{D_{\rm coh}}{v_\perp}
    \\
    &\simeq
    5.77\,\mathrm d
    \left(
        \frac{D_{\rm coh}}{0.1\,\mathrm{AU}}
    \right)
    \left(
        \frac{30\,\mathrm{km\,s^{-1}}}{v_\perp}
    \right).
\end{aligned}
    \label{eq:crossing_time_number}
\end{equation}
For the reference values
$D_{\rm coh}=0.1\,\mathrm{AU}$ and
$v_\perp=30\,\mathrm{km\,s^{-1}}$,
Eq.~(\ref{eq:crossing_time_number}) gives
$T_{\rm cross}=5.77\,\mathrm d$. We use this as the reference crossing
duration for that specific normalization. Equation~(\ref{eq:crossing_time_number})
is intended for predominantly transverse passages. For an axial or
wake-skimming arc, longitudinal evolution of the wake and rotation of the
baseline can also end the coherent interval, and the full projected waveform
must be used.

A $19\,\mathrm d$ interval does not correspond to the same pair of
parameters. It requires, for example,
$v_\perp=9.11\,\mathrm{km\,s^{-1}}$ at
$D_{\rm coh}=0.1\,\mathrm{AU}$, or
$D_{\rm coh}=0.329\,\mathrm{AU}$ at
$v_\perp=30\,\mathrm{km\,s^{-1}}$. A $19\,\mathrm d$ result should be quoted only with the corresponding values
of $D_{\rm coh}$ and $v_\perp$. More generally, a long coherent interval may
come from a smaller transverse speed, a larger effective coherence width, or a
combination of the two. The propagated trajectory and projected wake-Hessian
waveform must determine the duration; it cannot be assumed independently.

Third, the local tidal expansion requires the baseline to be short
compared with the spatial variation scale of the wake Hessian. The
corresponding condition is given in
Eqs.~(\ref{eq:hessian_variation_scale})
and~(\ref{eq:baseline_requirement}). When this condition is not
satisfied, the exact finite difference in
Eq.~(\ref{eq:exact_wake_difference}) must be used.

For a precession search, Eq.~(\ref{eq:tide_scaling}) gives
\begin{equation}
    M_{\rm od}^{\rm min}
    \sim
    \Msun
    \frac{\sigma_{\dot\omega}}{n}
    \left(\frac{R}{a}\right)^3
    \frac{1}{|\mathcal F|}.
    \label{eq:mmin_precession}
\end{equation}
For a fly-through search, Eq.~(\ref{eq:impulse}) gives
\begin{equation}
    M_{\rm od}^{\rm min}
    \sim
    \frac{v_{\rm rel}^2 b_{\rm fly}}{2G}\,
    \sigma_{\Delta\psi}.
    \label{eq:mmin_flyby}
\end{equation}
The compact and extended limits can be connected through the integrated
excess mass of a finite profile. For the Gaussian toy model in
Eq.~(\ref{eq:gaussian_wake}),
\begin{equation}
    M_{\rm od}^{\rm toy}
    =
    (2\pi)^{3/2}
    \delta\rho_{\rm pk}
    w^2\ell .
    \label{eq:gaussian_mass}
\end{equation}
For the normalization
$\delta\rho_{\rm pk}=0.4\,\mathrm{GeV\,cm^{-3}}$,
$w=1\,\mathrm{AU}$, and $\ell=1\,\mathrm{AU}$,
\begin{equation}
    M_{\rm od}^{\rm toy}
    \simeq
    1.9\times10^{-17}\,M_\odot .
    \label{eq:gaussian_mass_number}
\end{equation}
Equations~(\ref{eq:mmin_precession}) and (\ref{eq:mmin_flyby}) summarize the compact-source scalings. Equation~(\ref{eq:rho_min}) defines an equivalent uniform-core normalization. The mission reach requires a joint orbit and range fit with the physical wake template and nuisance parameters for known forces.

\subsection{Candidate low-transverse-speed trajectories}
\label{sec:best_trajectory}

Trajectory selection must ultimately be based on the post-fit signal-to-noise ratio of a propagated spacecraft formation. For illustration, the raw short-arc response may be written as
\begin{equation}
    \Delta L_{\parallel,{\rm raw}}
    \simeq
    -\frac{1}{2}\,
    L\,
    \hat L_i
    \mathcal E^w_{ij}(\mathbf r_g)
    \hat L_j
    T_{\rm coh}^2.
    \label{eq:trajectory_range_signal}
\end{equation}
A low-$v_\perp$ wake-skimming arc may increase the coherent response, but its value must be tested against transfer feasibility and baseline evolution. Non-gravitational acceleration noise and link geometry also matter, as do stationkeeping and covariance with fitted spacecraft states. 

We write the guiding-centre trajectory as
\begin{equation}
    \vecr_g(t)=s(t)\hatw+\mathbf b_\perp(t),
    \qquad
    \mathbf b_\perp\cdot\hatw=0 ,
    \label{eq:guide_centre_trajectory}
\end{equation}
and define the transverse speed through the wake tube,
\begin{equation}
    v_\perp
    \equiv
    \left|
    \dot{\vecr}_g
    -
    (\dot{\vecr}_g\cdot\hatw)\hatw
    \right| .
\end{equation}
The physical wake width $w_{\rm phys}(s)$ is obtained from the
calculated density profile and is estimated in
Eq.~(\ref{eq:wake_width}). Navigation and wake-direction uncertainties
are kept separate. Their transverse contribution is
\begin{equation}
    \sigma_b^2(s)
    =
    \sigma_{\rm nav}^2
    +
    s^2\sigma_\theta^2,
    \label{eq:trajectory_targeting_uncertainty}
\end{equation}
where $\sigma_{\rm nav}$ is the transverse navigation uncertainty and
$\sigma_\theta$ is the uncertainty in the adopted wake direction.

A robust targeting and dwell-time requirement may be written as
\begin{equation}
    b_\ast
    +
    N_\sigma\sigma_b
    \lesssim
    w_{\rm phys},
    \qquad
    v_\perp
    \lesssim
    \frac{D_{\rm coh}}{T_{\rm coh}},
    \label{eq:trajectory_requirements}
\end{equation}
where $N_\sigma$ specifies the desired targeting confidence and
$D_{\rm coh}$ is the effective transverse coherence width measured from
the projected tidal waveform. For an axial arc, the second inequality is only a necessary geometric
condition; by itself, it does not ensure coherence. Numerically,
\begin{equation}
    v_\perp
    \lesssim
    0.47\,\mathrm{km\,s^{-1}}
    \left(
        \frac{D_{\rm coh}}
        {0.1\,\mathrm{AU}}
    \right)
    \left(
        \frac{1\,\mathrm{yr}}
        {T_{\rm coh}}
    \right).
    \label{eq:vperp_requirement}
\end{equation}
This condition is an important input to the orbit design. For the
specific reference values
$D_{\rm coh}=0.1\,\mathrm{AU}$ and
$v_\perp=30\,\mathrm{km\,s^{-1}}$, the crossing lasts
$5.77\,\mathrm d$, which is shorter than one week. A
$19\,\mathrm d$ encounter requires the modified speed or coherence width
given below Eq.~(\ref{eq:crossing_time_number}). Relative to a one-year interval, the raw $T^2$ normalization is reduced by
\begin{equation}
    \left(
        \frac{5.77\,\mathrm d}{1\,\mathrm{yr}}
    \right)^2
    \simeq
    2.5\times10^{-4}
\end{equation}
for the reference crossing. For a $19\,\mathrm d$ crossing, the corresponding
factor is
\begin{equation}
    \left(
        \frac{19\,\mathrm d}{1\,\mathrm{yr}}
    \right)^2
    \simeq
    2.7\times10^{-3}.
\end{equation}
A low-$v_\perp$ wake-skimming arc remains a candidate for further study. Its optimality must be established by the full fit.

A candidate heliocentric orbital plane contains the nominal wake axis
only if its unit angular-momentum vector satisfies
\begin{equation}
    \hat{\mathbf h}
    (i_{\rm sci},\Omega_{\rm sci})
    \cdot
    \hat{\mathbf w}
    =
    0.
    \label{eq:orbit_plane_condition}
\end{equation}
This condition does not uniquely determine the inclination or node. It also
does not ensure a small transverse speed or a long dwell time. Transfer and
formation feasibility remain separate constraints. A propagated trajectory
calculation must determine the inclination, node, line of apsides, and science
interval.

A passage near apoapsis may reduce the total orbital speed, but
apoapsis alone does not guarantee a small velocity transverse to the
wake axis. Deterministic trim manoeuvres may be considered before the
science interval. Any manoeuvre within the fitted interval must be
included explicitly in the measurement model.

The preferred heliocentric distance is set by a compromise. Moving closer to
the Sun increases the focusing scale through $v_e^2(r)=2G\Msun/r$, but it
also raises solar radiation pressure and the orbital speed. Thermal recoil and
plasma noise in radio links become more demanding as well. Moving too far
outward improves dwell time but weakens the solar focusing and degrades
tracking leverage. For wave dark matter the natural target radius is the
wave-optics crossover,
\begin{equation}
    m\widetilde u(r_\ast) r_\ast/\hbar \sim 1 ,
\end{equation}
where $\widetilde u(r)$ is the characteristic local dark-matter speed
in the solar potential, defined in Eq.~(\ref{eq:focused_mode_speed}).
The same condition may be written as
\begin{equation}
    r_\ast
    \simeq
    1.6\,{\rm AU}
    \left(\frac{10^{-15}\,{\rm eV}}{m}\right)
    \left(\frac{240\,{\rm km\,s^{-1}}}{\widetilde u(r_\ast)}\right).
    \label{eq:trajectory_rstar}
\end{equation}
For $m\sim10^{-15}\,\mathrm{eV}$ and the reference speed, the wave-optics crossover occurs at a heliocentric distance of order $1$--$2\,\mathrm{AU}$. The preferred science radius must also reflect the focused contrast and physical wake width. Transfer and formation dynamics set additional constraints, together with acceleration noise, tracking geometry, and the asteroid environment.

For $m\sim10^{-16}\,\mathrm{eV}$, the same scaling gives a nominal
crossover distance of order $16\,\mathrm{AU}$ at the reference speed.
A mission at several AU would still lie in the wave-dominated regime, well
inside the nominal crossover radius.

The standard-halo template is the least structured benchmark considered
here. Its broad solar-frame velocity distribution superposes focused
contributions with different incident directions and focusing strengths,
leaving only a low-contrast net pattern near the terrestrial region
\cite{Kim:2021yyo}. An additional component occupying a narrower region of
velocity space would change both the amplitude and the target geometry. A
small solar-frame lag increases the deflection, while a small directional
spread prevents averaging over many wake axes. A cold stream tends
to define a compact, well-localized tube. A low-lag disk-like population can
produce a much broader feature. The existence and properties of either
component are astrophysical inputs, not assumptions imposed by the experiment.

For wave dark matter, kinematic broadening competes with diffraction. A useful
scaling for the angular support is \cite{Kim:2021yyo}
\begin{equation}
\begin{aligned}
    \Delta\mu_{\rm wave}
    &\sim
    \max\left[
        \frac{\hbar}{m\widetilde u(r)r},
        \left(
            \frac{\sigma}{v_\infty}
        \right)^2
    \right]
    \\
    &=
    \max\left[
        \frac{\ell_{\rm dB}}{r},
        \left(
            \frac{\sigma}{v_\infty}
        \right)^2
    \right].
\end{aligned}
    \label{eq:angular_width_wave}
\end{equation}
The first term is the diffraction floor, while the second is the angular
spread inherited from the incident velocity distribution. In the particle
limit only the latter remains,
$\Delta\mu_{\rm part}\sim(\sigma/v_\infty)^2$. This allows a spacecraft to
sample inclined stream wakes that Earth intersects weakly or not at all. Such
inclinations are useful discriminants because they move the target away from
many ecliptic-plane degeneracies. Since most planets and main-belt asteroids
remain close to the ecliptic, an out-of-plane science arc may also reduce the
strength and frequency of their perturbations, especially close encounters.
Their long-range tides and ephemeris errors still require explicit modelling.

The preferred baseline direction must be determined from the calculated
finite-profile Hessian. At a fixed position, the unconstrained
projected response is extremized along an eigenvector of
$\mathcal E^w_{ij}$, but that eigenvector need not be transverse to the
wake axis. Formation dynamics and link acquisition may also prevent the baseline from
following the instantaneous maximum-response direction. Attitude and
navigation constraints add further restrictions.

The local approximation requires the baseline to be short compared with
the spatial variation scale of the wake Hessian. Define
\begin{equation}
    \ell_{\mathcal E}^{-1}
    \equiv
    \frac{
        \left\|
            \nabla\mathcal E^w
        \right\|
    }{
        \left\|
            \mathcal E^w
        \right\|
    }.
    \label{eq:hessian_variation_scale}
\end{equation}
The leading-gradient approximation requires
\begin{equation}
    L\ll\ell_{\mathcal E}.
    \label{eq:baseline_requirement}
\end{equation}
A LISA-like value $L=0.0167\,\mathrm{AU}$ can be retained as a reference baseline. The exact finite difference in Eq.~(\ref{eq:exact_wake_difference}) should be used when $L/\ell_{\mathcal E}$ is not asymptotically small. Baseline rotation, repeated passages with different orientations, or additional probes would provide multiple tensor projections and help distinguish a wake from spacecraft-force and asteroid perturbations.

A mission study should compare several explicitly propagated cases with one
common force, noise, and covariance model. The set should contain the reference
crossing with $D_{\rm coh}=0.1\,\mathrm{AU}$ and
$v_\perp=30\,\mathrm{km\,s^{-1}}$. It should also contain one or more
$19\,\mathrm d$ cases with their corresponding $(D_{\rm coh},v_\perp)$ values,
plus candidate long-dwell trajectories.

For each dark-matter phase-space model, the calculation should provide the
downstream direction and physical density profile. It should then derive the
projected wake-Hessian waveform and targeting uncertainty. For each spacecraft
case, the study should report the transfer trajectory and formation initial
conditions. The report should also give the impact-parameter history and transverse
speed, together with the baseline evolution. Manoeuvre history and tracking
geometry should be stated separately. The final product is the post-fit wake
covariance.

The preferred configuration should be selected using the recovered
signal-to-noise ratio and parameter identifiability. Closest approach and raw
dwell time alone are insufficient. A wake-skimming
formation remains a scientifically interesting possibility because the
signal is tied to a Galactic flow direction and changes predictably
under baseline rotation or repeated sampling. Demonstrating that it is
the most sensitive configuration, however, requires the propagated
trajectory and covariance comparison described above.

\section{Signal enhancement from long-dwell trajectories and a low-lag dark disk}
\label{sec:long_dwell_dark_disk}

There are two direct ways to
increase the raw signal: remain in a coherent projected tide for longer,
and target a component with a larger focusing-induced excess density.
Equations~(\ref{eq:wave_delta_r_number}) and
(\ref{eq:percent_excess_range_number}) show the first effect through the
$T^2$ free-response scaling. To state its domain of validity, define
\begin{equation}
    \mathcal E_\parallel^w(t)
    \equiv
    \hat L_i(t)
    \mathcal E^w_{ij}\!\left[\mathbf r_g(t)\right]
    \hat L_j(t),
\end{equation}
and the dimensionless instantaneous projection
\begin{equation}
    \mathcal P_w(t)
    \equiv
    \frac{3\mathcal E_\parallel^w(t)}
    {4\pi G\,\delta\rho_{\rm pk}}.
    \label{eq:wake_projection_factor}
\end{equation}
For a reference baseline $L_\star$, introduce the response-weighted
projection
\begin{equation}
\begin{aligned}
    \overline{\mathcal P}_T
    \equiv
    \frac{2}{L_\star T^2}
    \int_{t_0}^{t_0+T}
    &(t_0+T-t)\,
    L(t)\,
    \mathcal P_w(t)\,
    \mathrm dt .
\end{aligned}
    \label{eq:weighted_projection_factor}
\end{equation}
The local-Hessian range response may then be written as
\begin{equation}
    \left|\Delta L_{\rm raw}\right|
    =
    \frac{2\pi G\,\delta\rho_{\rm pk}}{3}
    L_\star T^2
    \left|\overline{\mathcal P}_T\right|.
    \label{eq:coherent_range_generalized}
\end{equation}
A constant uniform core with a fixed baseline has
$\overline{\mathcal P}_T=1$. For a physical wake, $\overline{\mathcal P}_T$ tracks the changing wake
amplitude and baseline geometry. It also captures cancellation between
regions in which the projected Hessian has opposite signs. When the
finite-baseline expansion is not valid, the same quantity should be
constructed directly from Eq.~(\ref{eq:exact_wake_difference}).

For a genuinely coherent one-year interval, the raw normalization in Eq.~(\ref{eq:percent_excess_range_number}) becomes
\begin{equation}
\begin{aligned}
    \left|\Delta L_{\rm raw}\right|
    \simeq{}&
    2.5\,\mathrm{nm}
    \left(
        \frac{\delta\rho_{\rm pk}}
        {4.0\times10^{-3}\,\mathrm{GeV\,cm^{-3}}}
    \right)
    \left(
        \frac{L_\star}{0.0167\,\mathrm{AU}}
    \right)
    \\
    &\times
    \left(
        \frac{T}{1\,\mathrm{yr}}
    \right)^2
    \left|\overline{\mathcal P}_T\right|.
\end{aligned}
    \label{eq:one_year_raw_envelope}
\end{equation}
This is larger than the $5.77\,\mathrm d$ normalization by approximately $4.0\times10^3$. The significance of this scaling is simple: a geometry that holds the formation in the wake can move the raw response from below a picometre to the nanometre range. 

Extending the observing interval after a short crossing does not
preserve the quadratic scaling. If the spacecraft exits the wake at
$t_{\rm out}$, then at later times
\begin{equation}
\begin{aligned}
    \Delta L_{\rm raw}(T)
    &\simeq
    \Delta L_{\rm cross}
    +(T-t_{\rm out})\Delta v_w,
    \\
    \Delta v_w
    &\equiv
    \int_{\rm cross}
    \Delta a_\parallel^w(t)\,\mathrm dt .
\end{aligned}
    \label{eq:post_crossing_range}
\end{equation}
The post-crossing response grows at most linearly and can
covary strongly with the fitted relative velocity. Achieving the $T^2$
enhancement requires a trajectory that remains inside a region with a
coherent projected wake tide.

An approximately axial science arc is especially favorable. Writing
\begin{equation}
    \dot{\mathbf r}_g
    =
    v_\parallel\hat{\mathbf w}
    +\mathbf v_\perp,
    \qquad
    \mathbf v_\perp\cdot\hat{\mathbf w}=0,
\end{equation}
the spacecraft may move rapidly along the wake while maintaining a
small transverse impact-parameter drift. For a passive Keplerian orbit
whose plane contains the wake axis, $v_\perp\simeq h/r$, where $h$ is
the specific angular momentum. For an orbit with perihelion $q$ that
reaches the science radius $r$, the smallest $h$ occurs when $r$ is the
apoapsis. In that limiting ellipse,
$h_{\min}^2=2GM_\odot q r/(q+r)$, which approaches
$2GM_\odot q$ for $r\gg q$. Requiring $q\gtrsim R_\odot$
gives the approximate lower envelope
\begin{equation}
\begin{aligned}
    v_{\perp,\min}(r)
    &\sim
    \frac{\sqrt{2GM_\odot R_\odot}}{r}
    \\
    &\simeq
    2.9\,\mathrm{km\,s^{-1}}
    \left(
        \frac{1\,\mathrm{AU}}{r}
    \right).
\end{aligned}
    \label{eq:passive_minimum_vperp}
\end{equation}
This is an optimistic lower envelope: thermal and operational constraints
will generally require $q\gg R_\odot$ and hence a larger transverse speed.
For a narrow $D_{\rm coh}=0.1\,\mathrm{AU}$ feature, this corresponds
to
\begin{equation}
    T_{\rm coh}
    \lesssim
    60\,\mathrm d
    \left(
        \frac{D_{\rm coh}}{0.1\,\mathrm{AU}}
    \right)
    \left(
        \frac{r}{1\,\mathrm{AU}}
    \right),
    \label{eq:passive_dwell_envelope}
\end{equation}
up to the detailed orbit geometry. Thus a one-year dwell through a
$0.1\,\mathrm{AU}$-wide feature near $1\,\mathrm{AU}$ is not available on
this passive, solar-avoiding envelope. Even at the limiting value in
Eq.~(\ref{eq:passive_minimum_vperp}), one year would require a radius of
order $6\,\mathrm{AU}$. It becomes more plausible at several AU or for a
much broader focused component. Continuous thrust could reduce the
transverse drift, but it would also introduce a large low-frequency
force that must be calibrated. A passive geometry, with trim manoeuvres
outside the science interval, is the cleaner starting point.

A low-lag disk-like population provides a deliberately optimistic stress
test for the mission concept. Accreted co-rotating components can arise when
satellites are drawn toward the baryonic disk before disruption
\cite{Read:2008fh,2009MNRAS.397...44R}, while dissipative dark-sector models
can produce a thinner structure \cite{Fan:2013yva}. The latter class is more tightly restricted by stellar-kinematic analyses,
including Gaia results \cite{Schutz:2017tfp,Buch:2018qdr,2021A&A...653A..86W}.
We do not assume that either component is present. The benchmark isolates how
a small solar-frame lag and a broad focused footprint alter the ranging
response.

We separate the kinematic choice from the component density. For the former,
we adopt the illustrative solar-focusing point used in
Ref.~\cite{Kim:2021yyo},
\begin{equation}
\begin{aligned}
    \rho_{\rm comp}&=f_{\rm dd}\rho_{\rm ref},
    \\
    v_\infty&=50\,\mathrm{km\,s^{-1}},
    \qquad
    \sigma=50\,\mathrm{km\,s^{-1}}.
\end{aligned}
    \label{eq:dark_disk_benchmark}
\end{equation}
The separate choice $f_{\rm dd}=0.2$ is guided by simulations in which a
relatively quiescent Milky Way contains an accreted dark-disk fraction of
about $20\%$ or less~\cite{2009ApJ...703.2275P}. Their combination is an
optimistic normalization; a hotter
or less abundant component would give a smaller signal. The lag vector must
also be transformed into the adopted ecliptic frame before the trajectory is
designed.

Following Ref.~\cite{Kim:2021yyo}, the downstream particle estimate at $r\simeq1\,\mathrm{AU}$ gives
\begin{equation}
\begin{aligned}
    \delta_{w,\rm pk}^{\rm dd}
    &\sim
    \left(
        1+\frac{v_e^2}{\sigma^2}
    \right)^{1/2}-1
    \\
    &\simeq 0.3,
\end{aligned}
    \label{eq:dark_disk_contrast}
\end{equation}
which we use only to set a raw peak normalization. A particular formation
may sample a smaller contrast; in a wave template, the sign can also change. The same benchmark has no hierarchy between its dispersion
and mean lag. Formally,
\begin{equation}
    w_\sigma(s)
    \sim
    s\frac{\sigma}{v_\infty}
    \sim s,
    \label{eq:dark_disk_width}
\end{equation}
but here the small-angle estimate should be read only as evidence for an
order-unity angular footprint, not as a Gaussian width. Such a broad feature
could remain coherent over a distance of order $1\,\mathrm{AU}$. In that illustrative case, a one-year encounter would require
$v_\perp\lesssim4.7\,\mathrm{km\,s^{-1}}$, compared with
$0.47\,\mathrm{km\,s^{-1}}$ for the $0.1\,\mathrm{AU}$ reference feature.
This is comparable to the idealized passive envelope near $1\,\mathrm{AU}$,
so retargeting may be more important than active transverse-speed control.
The actual coherence width must still be extracted from the projected
Hessian, not inferred from $w_\sigma$ alone.

Taking $f_{\rm dd}=0.2$ and
$\delta_{w,\rm pk}^{\rm dd}=0.30$ gives
\begin{equation}
\begin{aligned}
    \delta\rho_{\rm pk}^{\rm dd}
    &=
    f_{\rm dd}\rho_{\rm ref}
    \delta_{w,\rm pk}^{\rm dd}
    \\
    &\simeq
    2.4\times10^{-2}\,
    \mathrm{GeV\,cm^{-3}}.
\end{aligned}
    \label{eq:dark_disk_excess_density}
\end{equation}
The corresponding uniform-core raw normalization is
\begin{equation}
\begin{aligned}
    \left|\Delta L_{\rm raw}^{\rm dd}\right|
    \simeq{}&
    3.8\,\mathrm{pm}
    \left(
        \frac{f_{\rm dd}}{0.2}
    \right)
    \left(
        \frac{\delta_{w,\rm pk}^{\rm dd}}{0.30}
    \right)
    \left(
        \frac{L_\star}{0.0167\,\mathrm{AU}}
    \right)
    \\
    &\times
    \left(
        \frac{T}{5.77\,\mathrm d}
    \right)^2
    \left|\overline{\mathcal P}_T\right|.
\end{aligned}
    \label{eq:dark_disk_range_number}
\end{equation}
In the idealized one-year coherent limit, the same expression becomes
\begin{equation}
\begin{aligned}
    \left|\Delta L_{\rm raw}^{\rm dd}\right|
    \simeq{}&
    15\,\mathrm{nm}
    \left(
        \frac{f_{\rm dd}}{0.2}
    \right)
    \left(
        \frac{\delta_{w,\rm pk}^{\rm dd}}{0.30}
    \right)
    \left(
        \frac{L_\star}{0.0167\,\mathrm{AU}}
    \right)
    \\
    &\times
    \left(
        \frac{T}{1\,\mathrm{yr}}
    \right)^2
    \left|\overline{\mathcal P}_T\right|.
\end{aligned}
    \label{eq:dark_disk_one_year_number}
\end{equation}
For this benchmark, the small mean speed and dispersion strengthen the
downstream contrast. At the same time, $\sigma/v_\infty\simeq1$ spreads the
contrast over a large angle. The stronger contrast raises the raw amplitude,
and the broad footprint relaxes the transverse-speed requirement for a long
encounter. A broad, slowly varying waveform can still be more covariant with
fitted spacecraft states and the Galactic tide. Greater width does not
guarantee easier recovery.

A second gravitational effect must also be kept in the orbit model. Before
solar focusing is applied, the Galactic dark disk has its own smooth
potential, $\Phi_{\rm dd}^{\rm gal}$. The Sun and spacecraft share the uniform
part of its acceleration, but not the local tide. Expanding about the Sun's
Galactic position $\mathbf R_\odot$ gives
\begin{equation}
\begin{aligned}
    A_{{\rm dd},i}^{\rm rel}(\mathbf r)
    &\equiv
    -\partial_i\Phi_{\rm dd}^{\rm gal}
    (\mathbf R_\odot+\mathbf r)
    +\partial_i\Phi_{\rm dd}^{\rm gal}(\mathbf R_\odot)
    \\
    &\simeq
    -\mathcal E_{ij}^{\rm dd,gal}r_j,
\end{aligned}
    \label{eq:dark_disk_galactic_tide}
\end{equation}
where
$\mathcal E_{ij}^{\rm dd,gal}\equiv
\partial_i\partial_j\Phi_{\rm dd}^{\rm gal}(\mathbf R_\odot)$.
For a spacecraft pair, this produces
\begin{equation}
    \Delta a_\parallel^{\rm dd,gal}
    \simeq
    -L\,
    \hat L_i
    \mathcal E_{ij}^{\rm dd,gal}
    \hat L_j.
    \label{eq:dark_disk_galactic_pair}
\end{equation}
This broad, slowly varying contribution enters both the guiding-centre
and relative motion. For $f_{\rm dd}=0.2$, its local density is
$\rho_{\rm dd}^{\rm gal}=0.08\,\mathrm{GeV\,cm^{-3}}$, and Poisson's
equation gives the characteristic trace
$\mathcal E_{ii}^{\rm dd,gal}=4\pi G\rho_{\rm dd}^{\rm gal}
\simeq1.2\times10^{-31}\,\mathrm{s^{-2}}$. A simple one-year
free-response estimate is of order $10\,\mu\mathrm m$ at a heliocentric
scale of $1\,\mathrm{AU}$, far too small to alter the transfer or
wake-targeting geometry. Across the reference inter-spacecraft baseline,
however, the corresponding raw scale is of order $10^2\,\mathrm{nm}$.
It must be retained as a low-frequency nuisance in a
nanometre-level range fit, even though it does not require a new class of
science trajectory.

A self-consistent benchmark should use the same local disk density and spatial
profile in the smooth Galactic potential and in the asymptotic component that
is solar focused. The lag vector fixes the wake direction. Its magnitude and
the velocity dispersion control the focusing, but none of these quantities
alone determines the Galactic Hessian. The unfocused field belongs in the
nominal trajectory model. The additional solar-focused source is
$\delta\rho_w=\rho_{\rm comp}\delta_w$; keeping the two terms separate avoids
double counting.

The final trajectory and covariance study must use the full particle or wave
density map. Equations~(\ref{eq:dark_disk_contrast}) and
(\ref{eq:dark_disk_width}) are only scalar summaries. The comparison should
start with the reference narrow crossing and a physical standard-halo
template. It should then test a long-dwell axial arc and the low-lag dark-disk
benchmark, with every case propagated through the same force and noise model.

\section{Numerical validation of leading response scalings}
\label{sec:numerical_validation}

We use numerical integrations to test the linear response of the implementation to an injected compact mass and an injected extended excess density. The simulations are performed using the \texttt{REBOUND} code \cite{2012A&A...537A.128R} with the \texttt{IAS15} integrator \cite{2015MNRAS.446.1424R}. Extra forces from the dark matter distributions are implemented with the \texttt{REBOUNDx} library \cite{2020MNRAS.491.2885T}. 

\subsection{Compact-source precession map}

\begin{figure}[h]
    \centering
    \includegraphics[width=\columnwidth]{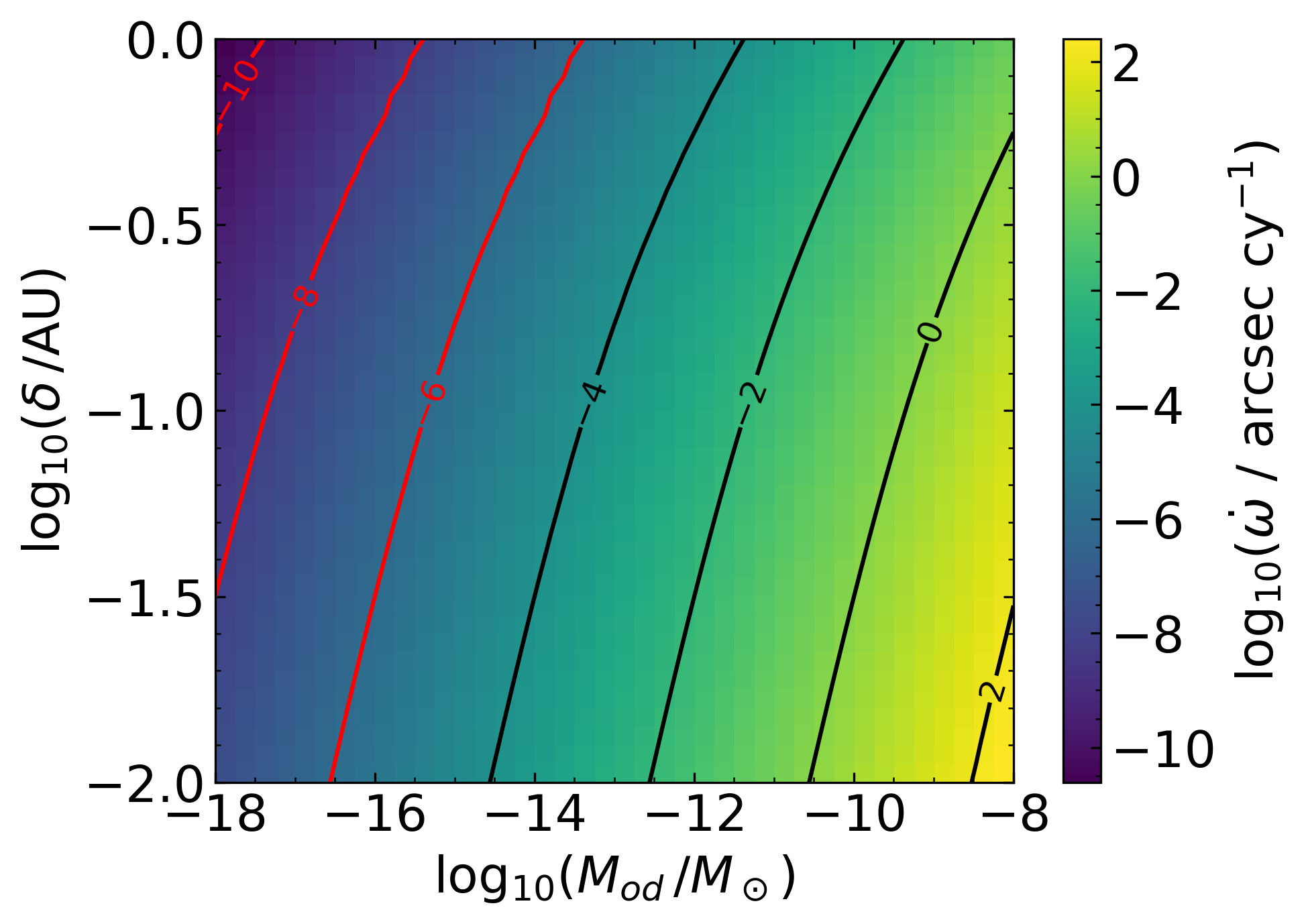}
    \caption{Numerical validation 1: precession rate of test-particle orbits about the Sun with $e=0.1$ in the presence of an external point mass centered at $x=0.8\,{\rm AU}$. The horizontal axis is $\log_{10}(M_{\rm od}/M_\odot)$, where $M_{\rm od}$ is the mass sourcing the point potential. The vertical axis is $\log_{10}(\delta/{\rm AU})$, where $\delta$ is the distance between the orbit pericenter and the perturber. Contours show $\log_{10}(\dot\omega/[{\rm arcsec}\,{\rm cy}^{-1}])$. Precession rates below $10^{-6}\,{\rm arcsec \ cy^{-1}}$ have been extrapolated using the linear dependence $\dot\omega\propto M_{\rm ext}$.
    }
    \label{fig:pre_grid_extrapolated}
\end{figure}

The first validation corresponds to the compact-source limit of Sec.~\ref{sec:signals}. Test-particle orbits with eccentricity $e=0.1$ are integrated around the Sun in an external point-mass potential centered at $x=0.8\,{\rm AU}$. The grid in Fig.~\ref{fig:pre_grid_extrapolated} varies both the source mass $M_{\rm od}$ and the pericenter separation $\delta$. The resulting map confirms the perturbative scaling $\dot\omega\propto M_{\rm od}$. It also shows the expected enhancement as the orbit approaches the perturber. This is the numerical counterpart of Eq.~(\ref{eq:tide_scaling}); order-unity corrections appear when the perturber is outside the asymptotic far-field regime.

\subsection{Two-spacecraft range residual}

\begin{figure}[h]
    \centering
    \includegraphics[width=\columnwidth]{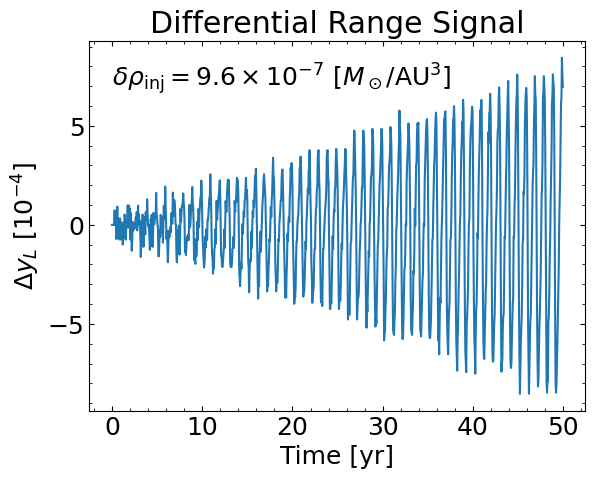}
    \caption{Numerical validation 2: fractional relative-range difference, $\Delta y_L=[L_{\rm inj}(t)-L_{\rm control}(t)]/L_{\rm control}(t)$, between two spacecraft initially separated by $0.0167\,\mathrm{AU}$. The injected source is a solid-sphere toy excess centered at $x=0.2\,\mathrm{AU}$ with radius $1\,\mathrm{AU}$. The injected excess density used in the plotted run is $\delta\rho_{\rm inj} =9.6\times10^{-7}M_\odot\,\mathrm{AU}^{-3}$.}
    \label{fig:strain_two_spacecraft}
\end{figure}

The second check probes the extended-density limit relevant for differential ranging. Two spacecraft are initialized with separation $L=0.0167\,{\rm AU}$ and propagated through the field of an offset solid-sphere overdensity. A control run removes the overdensity while keeping the remaining setup fixed. Figure~\ref{fig:strain_two_spacecraft} shows the resulting relative range residual,
\begin{equation}
    \Delta y_L(t)
    \equiv
    \frac{
        L_{\rm inj}(t)-L_{\rm control}(t)
    }{
        L_{\rm control}(t)
    }.
    \label{eq:numerical_fractional_range}
\end{equation}
The oscillatory residual provides a controlled demonstration that a finite excess density produces a coherent differential range signal after subtraction of the no-overdensity control run. The simulation supports the projected tidal observable in Eq.~(\ref{eq:projected_tidal_signal}) as the local quantity to fit in a full orbit-determination analysis.

\subsection{Density scaling of the differential signal}

The third calculation validate the linearity of the numerical response in the same extended-source setup. Over the directly simulated range, the fitted response amplitude is approximately proportional to the injected excess density, as expected in first-order perturbation theory. At smaller injections, the control-subtracted result falls below the numerical resolution of the present implementation.

This linearity checks the implementation. The static solid-sphere yearly-envelope fit and the short-arc raw uniform-core normalization in
Eq.~(\ref{eq:wave_delta_r_number}) use different source geometries, time dependences, and fitting conventions; the latter is evaluated at $T=5.77\,\mathrm d$.

\begin{figure}[h]
    \centering
    \includegraphics[width=\columnwidth]{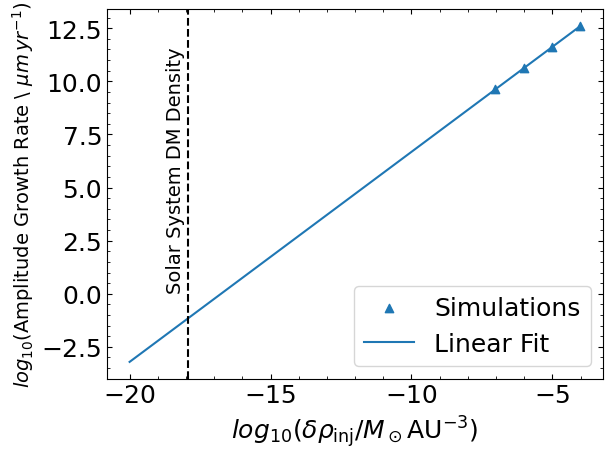}
    \caption{Numerical validation 3: fitted growth parameter of the yearly range-residual oscillation as a function of the injected toy excess density. The points show the directly simulated cases and the line shows a linear fit. The approximate linearity over the simulated range is a code validation of the first-order density dependence.}
    \label{fig:amplitude_fits}
\end{figure}

Together, these calculations check the sign and linear amplitude dependence of the simplified numerical implementation. A physical forecast requires injection and recovery of a computed particle or wave wake along a propagated spacecraft formation, with Solar-System and spacecraft nuisance parameters included in the fit.

\section{Backgrounds and discriminants}
\label{sec:backgrounds}

A search for a solar dark-matter wake is a precision orbit-determination problem. A residual left after subtracting a fixed Solar-System model is not enough. The wake template must improve a global fit in which the gravitational model, ephemeris parameters, and spacecraft forces vary simultaneously. This is the same logic that should be adopted in planetary-ephemeris limits on smooth Solar-System dark matter \cite{Pitjev:2013sfa}, asteroid-astrometry searches for fifth forces \cite{Tsai:2021irw}, and the Bennu/OSIRIS-REx fifth-force analysis \cite{Tsai:2023zza}. We discuss relativistic and solar terms first, followed by planetary and small-body gravity. Spacecraft forces and measurement errors form the remaining background classes.

\paragraph{Relativistic and solar terms.}
The leading post-Newtonian precession around the Sun is
\begin{equation}
    \dot\omega_{\rm GR}
    =
    \frac{3nG\Msun}{a c^2(1-e^2)}.
    \label{eq:gr_precession}
\end{equation}
For a sufficiently precise trajectory this term is not an optional correction; it is part of the reference model. In a global fit it is accompanied by the solar gravitational parameter and possible solar mass loss. The post-Newtonian parameters and solar quadrupole moment must also be fitted. The solar oblateness gives
\begin{equation}
    \dot\omega_{J_{2\odot}}
    =
    \frac{3nJ_{2\odot}R_\odot^2}
    {4a^2(1-e^2)^2}
    \left(5\cos^2 i_\odot-1\right),
    \label{eq:j2_precession}
\end{equation}
where $i_\odot$ is measured relative to the solar equator. This is an important covariance direction because it produces a secular perihelion effect, as do many smooth-density or long-range-force signals. Asteroid radar studies have emphasized this covariance between post-Newtonian parameters and $J_{2\odot}$ \cite{Verma:2017ywb}. A directed wake has a different structure. It is anisotropic and confined to a finite downstream region; in a two-spacecraft configuration, it appears as a projected tidal tensor.

\paragraph{Planetary and small-body gravity.}
Let $\mathbf r$ be the heliocentric spacecraft position and let $\mathbf r_X(t)$ be the heliocentric position of a Solar-System perturber $X$.  The heliocentric acceleration from that body is
\begin{equation}
    \mathbf a_X(\mathbf r,t)
    =
    GM_X
    \left[
    \frac{\mathbf r_X-\mathbf r}
    {|\mathbf r_X-\mathbf r|^3}
    -
    \frac{\mathbf r_X}{|\mathbf r_X|^3}
    \right].
    \label{eq:heliocentric_perturber_accel}
\end{equation}
The first term is the direct acceleration of the spacecraft; the second is the indirect acceleration of the heliocentric frame. The deterministic model contains the eight planets and the Moon. It also contains Pluto, the largest trans-Neptunian objects, and a substantial set of main-belt asteroids. Massive planets are not noise-like backgrounds: their phases and masses are tightly constrained, and their perturbations are highly predictable. Their initial conditions and gravitational parameters must nevertheless remain in the covariance matrix. Ephemeris-frame parameters must remain there as well, because errors in these quantities can project onto slow range and range-rate residuals.

The small-body contribution is more subtle.  Modern planetary ephemerides do not treat the asteroid belt as a single smooth object.  They include hundreds of individual asteroid perturbers and represent the unresolved remainder by ring or discrete-ring components \cite{2018AstL...44..554P,Park_2021}. This structure is essential for Gravity Probe--DM because a trajectory near the ecliptic can be more strongly affected by asteroid mass errors than by the nominal uncertainty in the major-planet ephemerides.  If the asteroid positions are known but their masses have errors $\delta M_A$, the corresponding acceleration error is
\begin{equation}
\begin{aligned}
\delta\mathbf a_{\rm ast}
=&\;
\sum_A G\,\delta M_A
\left[
\frac{\mathbf r_A-\mathbf r}{|\mathbf r_A-\mathbf r|^3}
-
\frac{\mathbf r_A}{|\mathbf r_A|^3}
\right]
\\
&+
\frac{\partial \mathbf a_{\rm ring}}{\partial M_{\rm ring}}\,
\delta M_{\rm ring}
+\cdots \, .
\end{aligned}
\label{eq:asteroid_mass_error_accel}
\end{equation}
For two spacecraft separated by $\mathbf L=L\hat{\mathbf L}$, the same mass
errors generate a differential acceleration
\begin{equation}
\begin{aligned}
\delta\Delta a_{\parallel}^{\rm ast}
\simeq\;&
-GL\,\hat L_i\hat L_j
\sum_A \delta M_A
\left(
\frac{\delta_{ij}}{d_A^3}
-
3\frac{d_{A,i}d_{A,j}}{d_A^5}
\right)
\\
&+\,
\delta\Delta a_{\parallel}^{\rm ring}
+\cdots ,
\qquad
\mathbf d_A\equiv\mathbf r_A-\mathbf r .
\end{aligned}
\label{eq:asteroid_tidal_error}
\end{equation}
Equation~(\ref{eq:asteroid_tidal_error}) shows why asteroid masses are a particularly important uncertainty for a gradiometer.  The error scales as $\delta M_A/d_A^3$ and can be dominated by a comparatively small
object during a close approach.  In the actual analysis one should not replace this by a white-noise floor.
The largest asteroid masses should enter as nuisance parameters. The fit should also include a main-belt ring model and the leading principal components of the asteroid-mass covariance. Close encounters with poorly measured small bodies should either be modelled explicitly or vetoed by trajectory design.

The trans-Neptunian population and Kuiper-belt ring are less important for short-period structure in the inner Solar System, but they generate slowly varying low-order multipoles. Their effect can covary with long-timescale drifts and solar $GM$, as well as with ephemeris-frame parameters. A mission extending to several AU should include Pluto and the largest trans-Neptunian objects. The reference model also needs a Kuiper-belt ring or an equivalent discrete representation.

\paragraph{Galactic and dark-disk tides.}
The Galactic field is nearly uniform across the Solar System, but a uniform acceleration is not the whole effect. Its local Hessian produces a slowly varying heliocentric tide. If a dark-disk benchmark is included,
Eqs.~(\ref{eq:dark_disk_galactic_tide}) and (\ref{eq:dark_disk_galactic_pair}) must be added to the propagated guiding-centre and relative motion. This term is broader than the solar-focused wake and follows the Galactic disk geometry. It does not look like a localized downstream crossing. Even so, it can covary with spacecraft initial conditions and ephemeris-frame parameters. Other low-frequency tides
create related degeneracies. The disk density and spatial profile should be constrained with the same stellar-kinematic priors used to define the benchmark \cite{Schutz:2017tfp,Buch:2018qdr,2021A&A...653A..86W}. Those quantities should be linked consistently between the smooth Galactic field and the focused component. The lag vector fixes the wake direction, and its
magnitude and the velocity dispersion enter the focusing calculation; they do not directly set the smooth Galactic Hessian.

\paragraph{Non-gravitational spacecraft forces.}
For a single spacecraft, non-gravitational accelerations can be a leading background. Solar radiation pressure and thermal recoil are central terms. Outgassing, leaks, and antenna recoil can add further forces, as can the solar wind when relevant. Attitude-control activity and deterministic manoeuvres must also be tracked. The force model usually contains an area-to-mass ratio and reflectivity coefficient. Attitude history and thermal time constants enter separately, together with impulsive or finite-burn parameters.
Schematically,
\begin{equation}
    \mathbf a_{\rm sc}
    =
    \mathbf a_{\rm SRP}
    (C_R,A/m,\mathbf r,\hat{\mathbf n})
    +
    \mathbf a_{\rm th}(\Theta_{\rm th})
    +
    \mathbf a_{\rm man}(\Theta_{\rm man})
    +\cdots .
    \label{eq:spacecraft_force_model}
\end{equation}
These terms have different symmetries from a gravitational wake. Solar radiation pressure follows the Sun--spacecraft direction and scales approximately as $r^{-2}$. Thermal recoil follows the spacecraft geometry and
thermal history. Manoeuvres are localized in time and correlated with telemetry. A two-spacecraft measurement rejects a spatially uniform acceleration, but it does not reject the solar or planetary gravity gradients across the baseline. Those tides remain in the differential measurement and must be modelled jointly with the wake and spacecraft forces. The mission needs stable drag-free or accelerometer-assisted operation. An alternative is a force model accurate enough to keep the residual differential acceleration below the projected wake signal.

\paragraph{Measurement and ephemeris errors.}
The primary observables are range, range rate, and Doppler. Clock comparison provides another channel; a two-spacecraft mission also supplies inter-spacecraft range or range acceleration. The nuisance model must include clock and ranging noise, along with station-location errors. Ground links add tropospheric and plasma delays. Time-scale conversion, antenna phase-centre
offsets, and ephemeris-frame rotations complete the measurement model. A convenient schematic likelihood is
\begin{equation}
    \mathbf y(t)
    =
    \mathbf y_{\rm SS}(t;\Theta_{\rm eph})
    +
    \mathbf y_{\rm sc}(t;\Theta_{\rm sc})
    +
    A_w\,\mathbf h_w(t;\Theta_w)
    +
    \mathbf n(t),
    \label{eq:background_likelihood}
\end{equation}
where $\Theta_{\rm eph}$ contains solar and planetary parameters, together with asteroid and trans-Neptunian parameters. The set $\Theta_{\rm sc}$ contains the spacecraft-force and instrument parameters, and $\mathbf h_w$ is the dark-matter wake template. The wake amplitude $A_w$ and shape parameters $\Theta_w$ should be fit at the same time as the nuisance parameters, not after subtracting a fixed ephemeris.

\paragraph{Discriminants of a focused dark-matter wake.}
Despite these backgrounds, a focused wake has several robust discriminants. Its predicted downstream direction satisfies $\hat{\mathbf w}\simeq\hat{\mathbf v}_\infty$. That direction is generally unrelated to the ecliptic, the Sun--spacecraft line, or the phase of a known asteroid. The apparent sky direction from which the flow arrives is $-\hat{\mathbf v}_\infty$. The signal is also localized: it turns on and off as the spacecraft enters and exits a finite wake tube. In a two-spacecraft configuration it produces a correlated tidal response,
\begin{equation}
    \Delta a_\parallel^w
    \simeq
    -L\,
    \hat L_i
    \mathcal E^w_{ij}
    \hat L_j ,
    \label{eq:dm_tidal_discriminant}
\end{equation}
whereas a uniform acceleration is common mode and cancels at leading order. The signal also changes predictably under baseline rotation and trajectory redesign. Planetary and asteroid perturbations follow known Solar-System orbits. Non-gravitational forces follow illumination, attitude, and manoeuvres. The wake instead follows the incident dark-matter flow.

Wave dark matter adds another discriminator. The de Broglie scale controls the wake smoothing, while the propagated phases set any temporal or spatial interference pattern; see Eqs.~(\ref{eq:debroglie})--(\ref{eq:x_wave_parameter}).
A particle-like caustic, a dispersion-broadened stream, and a wave-smoothed interference pattern produce different crossing profiles. The decisive test is an injection-and-recovery covariance study. It must fit the particle and wave wake templates together with Solar-System ephemeris and
asteroid-mass parameters. Unresolved-belt components and spacecraft-force nuisance parameters must be included in the same fit.

\section{Discussion and outlook}
\label{sec:conclusion}

We have proposed \emph{Gravity Probe--DM}: a heliocentric mission concept to search gravitationally for the solar dark-matter wake produced by solar focusing. The central point is that the Sun is not merely a source of background gravity. It is also a gravitational lens for any incident Galactic dark-matter flow. We call the resulting structure an irreducible target. Its amplitude and shape depend on the incident phase-space distribution. Wave interference may produce depleted as well as enhanced fringes. Even so, an unbound component cannot pass through the solar potential without being perturbed. Gravity Probe--DM turns that unavoidable gravitational imprint into a spacecraft observable.

Cold streams, debris flows, and dark disks can generate directional downstream structures. Wave dark matter can do so as well. The incident phase-space distribution fixes the position and width of the structure; for a wave component, it also fixes the coherence scale. A formation that follows the
wake for a year increases the raw one-percent normalization by about $4.0\times10^3$, from $0.63\,\mathrm{pm}$ to $2.5\,\mathrm{nm}$. For
the illustrative low-lag dark-disk benchmark, the same one-year envelope is about $15\,\mathrm{nm}$. Those pre-fit response scales show that trajectory design and the local phase-space structure can change the signal by orders of magnitude.

The mission concept is developed primarily to search for the
solar-focused wake, which poses a different question from existing Solar-System limits on smooth dark matter. Planetary and asteroid ephemerides already constrain approximately spherical excess-mass distributions around the Sun. A focused wake is instead anisotropic and localized, with its direction tied to a specific incident dark-matter flow. Its most
distinctive two-spacecraft observable is the trajectory-dependent projection of the wake Hessian measured through differential ranging.

The same heliocentric tracking architecture would remain sensitive to all three signal classes summarized in Sec.~\ref{sec:signal_classes}. A compact-object encounter would leave an acceleration transient in the individual tracking data and a tidal transient in the inter-spacecraft link \cite{Seto:2004zu,Tran:2023jci}. A smooth or slowly varying Solar-System component would produce a secular or long-duration perturbation, to be constrained through the global orbit solution \cite{Pitjev:2013sfa,Verma:2017ywb,Tsai:2021irw,Tsai:2022jnv}. These searches use different spatial templates and timescales, and each needs its own nuisance model. A trajectory designed for the solar-focused wake need not maximize sensitivity to the other two classes. The data could still support all three analyses, allowing Gravity Probe--DM to operate more broadly as a gravitational observatory for compact, smooth, and solar-focused dark structures in the Solar System.

The analytic estimates derived above summarize the physics of the proposal. In the compact-source limit, a localized excess mass produces a tidal precession
\begin{equation}
    \dot\omega
    \sim
    n\frac{M_{\rm od}}{M_\odot}
    \left(\frac{a}{R}\right)^3 ,
\end{equation}
where \(M_{\rm od}\) is the effective overdensity mass and \(R\) is its
distance from the Sun. The spacecraft orbit has semimajor axis \(a\) and mean
motion \(n\). A close crossing of a compact overdensity gives the impulse
estimate
\begin{equation}
    \Delta\psi
    \sim
    \frac{2GM_{\rm od}}
         {v_{\rm rel}^2 b_{\rm fly}} ,
\end{equation}
where \(b_{\rm fly}\) is the fly-through impact parameter. 
For an extended wake, a single short-baseline link measures a projected
component of the wake Hessian. In the uniform-core normalization, the raw free-response range scale is
\begin{equation}
    |\Delta L_{\rm raw}^{\rm core}|
    \simeq
    \frac{2\pi G\,\delta\rho_{\rm core}}{3}
    LT^2 .
    \label{eq:conclusion_deltar}
\end{equation}
Here $L$ is the inter-spacecraft baseline, $T$ is the coherent interval, and $\delta\rho_{\rm core}$ is the excess density of the normalization profile. Equation~(\ref{eq:conclusion_deltar}) gives the free response before
formation-state fitting and nuisance marginalization. This response is linear in the baseline and quadratic in the coherent interval. The recovered sensitivity can follow a different scaling. Slowly varying components can covary with the spacecraft initial states and Solar-System parameters. Low-frequency force noise and instrumental drifts create further covariance.

The long-dwell calculation nevertheless identifies a clear mission direction. The short transverse crossing should remain the conservative benchmark. An axial wake-skimming arc and a low-lag dark disk are the principal
enhancement cases. The axial case tests how much of the $T^2$ response survives realistic formation dynamics. The dark-disk case asks whether a stronger, broader source can sustain a long coherent waveform. It does not require a new class of heliocentric orbit, although its lag vector and focused profile may shift the preferred orbital plane and encounter epoch. Its smooth Galactic tide is too weak to drive mission design. The tide must still be propagated as a low-frequency nuisance, with its density and spatial profile linked consistently to the focused component.

A detection would have a particularly clean interpretation. It would be a purely gravitational observation of dark matter in the Solar System, independent of any assumed coupling to the Standard Model. It would also map the local distribution in the same environment occupied by terrestrial direct-detection experiments. A null result would still be valuable. It could exclude specified solar-focused streams and dark-disk components, as well as wave-interference patterns. The absence of a spacecraft residual would then become information about the local dark-matter phase-space distribution.

The next stage is an end-to-end covariance study. It requires ephemeris-level trajectories for the reference transverse crossing, one or more long-dwell axial arcs, and the low-lag dark-disk direction. Particle and wave wake templates should be computed from the formalism of Ref.~\cite{Kim:2021yyo}. The propagation must include both the smooth Galactic
tide and the solar-focused perturbation. It must produce range and range-rate observables, as well as Doppler and differential acceleration.

The fit must vary the wake amplitude together with solar $GM$ and solar $J_{2\odot}$. Post-Newtonian terms and planetary parameters must vary as well. Asteroid masses and unresolved-belt components belong in the same fit, along with the Galactic-tide parameters. The spacecraft model must add radiation pressure and thermal recoil. Manoeuvres, clock noise, and tracking systematics complete the model.

The numerical programme has four main milestones. First comes an injection-and-recovery test with planets and asteroids. A wave-focusing template study follows. A finite-baseline test must then compare the exact acceleration difference with the local scaling $\Delta L_{\rm raw}\propto L$ across a range of $L/\ell_{\mathcal E}$. Finally, the short-crossing, long-dwell, and dark-disk cases require a like-for-like covariance comparison.

Existing data can already help develop this analysis pipeline. LISA Pathfinder is the most relevant case, playing an methodological role. Its two test masses were separated by only
\begin{equation}
    \ell_{\rm LPF}\simeq 0.38\,{\rm m},
\end{equation}
so the direct tidal response to a broad solar dark-matter wake is extremely small. For the uniform-core normalization, the local wake-Hessian scale is
\begin{equation}
    |\mathcal E^w|
    \sim
    \frac{4\pi G\,\delta\rho_{\rm core}}{3}.
\end{equation}
The LPF differential-acceleration response is only
\begin{equation}
    \Delta a_{\rm LPF}^{w}
    \sim
    |\mathcal E^w|\ell_{\rm LPF}
    \simeq
    7.6\times10^{-32}\,\mathrm{m\,s^{-2}}
    \left(
        \frac{\delta\rho_{\rm core}}
        {0.4\,\mathrm{GeV\,cm^{-3}}}
    \right).
    \label{eq:lpf_dm_scale}
\end{equation}
This normalization already assumes a wake excess equal
to the full $0.4\,\mathrm{GeV\,cm^{-3}}$ reference density. A
one-percent wake excess gives a response smaller by another factor of $100$. Both are far below the published LPF differential-acceleration noise.
For the broad wake benchmark in Eq.~(\ref{eq:conclusion_deltar}), archival
LPF data support method development through flight data for drag-free control and optical metrology, together with measured disturbance channels and glitches. They also offer a mature framework for residual-force modelling
\cite{Armano:2016bkm,Armano:2018kix,LISAPathfinder:2024ucp,Hewitson:2009zz}.

A practical LPF reanalysis would begin with a dark-matter flow model $\Theta_w$. One would compute its solar-focused excess-density field $\delta\rho_w$ and solve for the wake Hessian $\mathcal E^w_{ij}$. The final step is to project that Hessian along the actual LPF ephemeris and attitude
history. The signal in the differential-acceleration channel would be
\begin{equation}
    h_{\rm LPF}(t;\Theta_w)
    =
    -\ell_{\rm LPF}\,
    \hat x_i(t)\,
    \mathcal E^w_{ij}
    \!\left[
        \mathbf r_{\rm LPF}(t);
        \Theta_w
    \right]\,
    \hat x_j(t),
    \label{eq:lpf_template}
\end{equation}
where \(\hat{\mathbf x}(t)\) is the sensitive axis joining the two test masses. The measured residual acceleration could then be modeled as
\begin{equation}
    d(t)
    =
    A_w h_{\rm LPF}(t;\Theta_w)
    +
    \sum_\alpha c_\alpha q_\alpha(t)
    +
    \sum_\beta g_\beta(t;\vartheta_\beta)
    +
    n(t),
    \label{eq:lpf_likelihood}
\end{equation}
where \(A_w\) is the wake amplitude. The channels \(q_\alpha(t)\) contain environmental and instrumental
measurements. They include temperature and magnetic-field monitors, together with charge and actuation channels. Spacecraft-control diagnostics enter here as well. The terms
\(g_\beta\) represent transients or glitch models, and \(n(t)\) is the stochastic residual. This follows the template-based anomalous-tidal-stress analyses previously discussed for LPF \cite{Korsakova:2014mya}. Here the tidal tensor is supplied by solar-focused dark matter, not by a modified-gravity model.

The LPF search should be reported as a validation exercise and a specialized null test. The relevant frequency content is determined by the full projected template $h_{\rm LPF}(t;\Theta_w)$ evaluated along the actual LPF ephemeris and
attitude history. For a simple transverse crossing through a spatial feature of wavelength $\lambda_{\rm sig}$,
\begin{equation}
    f_{\rm sig}
    \sim
    \frac{v_\perp}{\lambda_{\rm sig}}.
    \label{eq:lpf_signal_frequency}
\end{equation}
For a nonoscillatory envelope of path length $D_{\rm sig}$, the corresponding duration is $T_{\rm sig}\sim D_{\rm sig}/v_\perp$, with a broad frequency scale $f_{\rm env}\sim1/T_{\rm sig}$. Navigation and pointing uncertainty affect the predicted template but do not determine the physical signal frequency. A single width-based estimate is
insufficient. The final analysis should use the Fourier transform of the complete projected wake template.

Compact wakes and sharp stream features are better suited to an LPF-style matched-filter test, as are wave-interference templates with short spatial scales. Such an analysis would test the machinery needed for a future mission. This includes ephemeris-based template generation and projection onto an
instrument axis, followed by whitening with real nonstationary noise. It would also test glitch vetoes, Bayesian evidence or matched-filter statistics, and injection-and-recovery calibration.

Looking forward, a dedicated heliocentric gradiometer should be evaluated around the strongest scientifically allowed configuration, not only the simplest crossing. Baseline selection should compare the exact finite-difference response with the local Hessian-variation scale. Link and formation constraints must then be folded into the post-fit nuisance
covariance. The trajectory study should give equal weight to an axial long-dwell arc and a low-lag dark-disk wake, while retaining the standard halo as the irreducible reference target.

Baseline rotation, repeated passages, and additional probes would provide independent tensor projections. These measurements could separate a Galactic-flow signal from the smooth Galactic tide and from asteroid perturbations. Their distinct correlations would also help reject radiation pressure, thermal recoil, and clock-like range errors. In this form, Gravity Probe--DM would use the Solar System as a dark-matter lensing laboratory: the Sun focuses the dark matter, the spacecraft sample the resulting structure, and precision ranging converts its gravitational field into a measurable spatial template.

\subsection{Networked probes and differential observables}
\label{sec:probe_network}

A two-spacecraft formation is the minimal element of a more general
heliocentric gravity-gradient network. For probes $a$ and $b$ at positions
$\mathbf r_a$ and $\mathbf r_b$, define
\begin{equation}
    \mathbf L_{ab}
    \equiv
    \mathbf r_b-\mathbf r_a,
    \qquad
    L_{ab}=|\mathbf L_{ab}|,
    \qquad
    \hat{\mathbf L}_{ab}
    =
    \frac{\mathbf L_{ab}}{L_{ab}}.
\end{equation}
The exact wake contribution to the differential line-of-sight acceleration is
\begin{equation}
\begin{aligned}
    \Delta a^w_{ab,\parallel}
    =
    \hat{\mathbf L}_{ab}\cdot
    \left[
        -\nabla\Phi_w(\mathbf r_b)
        +\nabla\Phi_w(\mathbf r_a)
    \right].
    \label{eq:network_exact_difference}
\end{aligned}
\end{equation}
When the link is short compared with the physical variation scale of the wake,
this becomes
\begin{equation}
    \Delta a^w_{ab,\parallel}
    \simeq
    -L_{ab}\,
    \hat L_{ab,i}
    \mathcal E^w_{ij}(\mathbf r_{g,ab})
    \hat L_{ab,j},
    ~~
    \mathbf r_{g,ab}
    =
    \frac{\mathbf r_a+\mathbf r_b}{2}.
    \label{eq:network_local_difference}
\end{equation}
Each link measures one scalar projection of the local tidal tensor.
A single baseline cannot, by itself, determine the full tensor or separate all
wake-profile and trajectory parameters.

Additional probes do more than extend the integration time. Three probes in a
non-collinear formation provide simultaneous projections along three in-plane
baselines. A non-coplanar four-probe formation has six links and can, in
principle, constrain all six independent components of a symmetric tidal
tensor. The actual recovery remains subject to orbit-state covariance, force
uncertainties, and link noise. For a compact formation whose links sample
approximately the same point, recovery of all tensor components would also
allow the trace to be compared with
\begin{equation}
    \mathcal E^w_{ii}
    =
    4\pi G\,\delta\rho_w .
\end{equation}
In practice, Eq.~(\ref{eq:network_exact_difference}) should be used to evaluate finite-baseline corrections and field variation between link midpoints, as a spatially constant Hessian need not be adequate.

A complementary architecture uses separated spacecraft pairs. Different pairs could skim the wake at different impact parameters and heliocentric distances; others could sample distinct longitudinal positions. Together they would map the transverse width and longitudinal structure. They would also help separate the peak excess density from impact parameter and dwell time. One pair could remain near the predicted wake core while another follows an offset trajectory or samples a control region outside the nominal wake. Staggered crossings
would test whether one wake direction and profile reproduce every observed feature, including its sign, duration, and amplitude.

The network should still be analysed as a differential orbit-determination problem. Differencing does not remove solar, planetary, or asteroid tidal fields. They appear coherently in every link and must be propagated through the same model. The advantage is overconstraint: one common set of phase-space
and trajectory parameters must reproduce several baseline projections and spatial crossings. Solar-radiation-pressure and thermal-recoil errors have distinct correlations across the network. So do link-specific clock and ranging errors, as well as individual asteroid encounters. A joint fit can exploit these differences while retaining the shared ephemeris and force parameters in the covariance matrix.

A network also addresses the intrinsic degeneracies of a single passage. In one range waveform, wake density and width can trade against impact parameter. Transverse velocity, dwell time, and baseline projection create further trade-offs. Multiple link orientations constrain the tensor projection, while multiple impact parameters constrain the spatial profile. Independent navigation fixes the spacecraft geometry. A probe outside the main wake region then provides a control for long-timescale instrumental and ephemeris
drifts. These measurements would not guarantee a detection. They would turn a one-dimensional range anomaly into a spatially and tensorially constrained test of a solar-focused wake.

It is natural to consider plan a staged implementation. A two-probe mission would test the basic heliocentric differential-ranging concept. A three- or four-probe formation would add simultaneous tensor projections. Several separated pairs would instead enable wake tomography over a larger volume. Architecture selection should first account for trajectory feasibility and physical wake width. Link acquisition, low-frequency acceleration noise, and asteroid confusion set the main operational trade-offs. The final choice should be made
from the rank and uncertainty of the marginalized wake-parameter covariance.

The network would also broaden the mission beyond the wake search. Compact-object encounters could be tested through their timing and amplitude across several links; the baseline dependence supplies another check. A smooth Solar-System dark-matter component would produce a spatially coherent, slowly varying response across the network. These signals require analyses distinct from the directional wake template. Multiple probes would nonetheless provide valuable redundancy and help separate gravitational perturbations from disturbances associated with one spacecraft or link.

\subsection{Implications for dark-matter properties}
\label{sec:dm_property_implications}

Joint sampling of the wake at several heliocentric radii turns solar focusing into an inverse problem for the incident dark-matter component. In a particle model, the direction and width constrain the solar-frame flow. The tidal amplitude constrains the component density within the fitted geometry. A single phase-space model must reproduce the radial evolution of the signal. Coherent residual structure could reveal a superposition of incident components.

For wave dark matter, the radial evolution carries a mass scale. The local de Broglie length varies with the focused speed, so one value of $m$ must reproduce the smoothing and any sign changes along the trajectory \cite{Kim:2021yyo}. Measurements at multiple radii can separate diffraction from broadening caused by the incident velocity distribution and probe the particle-to-wave transition.

A gravitational reconstruction would supply a coupling-independent calibration of the Solar-System phase-space distribution. That distribution shapes the time and frequency structure of direct-detection signals \cite{Green:2017dd,Freese:2013modulation,Foster:2018halo}.
Combining the two measurements would help separate astrophysical uncertainty from interaction strength.

Mission results should therefore be reported in the physical parameter space of each wake template. A detection identifies the compatible region. A null result excludes the region accessible at the validated post-fit sensitivity. The equivalent uniform-core density can remain as a compact diagnostic. Ultimately, Gravity Probe--DM would use the Sun as
a lens and the wake as a gravitational fingerprint of the dark matter flowing through the Solar System.

\begin{acknowledgments}

We thank Scott Thomas for bringing the GRACE mission to our attention. We thank Erwin Tanin for useful discussions on dark-matter overdensities in the Solar System, and Asantha Cooray for valuable comments on the draft. Y.-D.T. is supported by a Dorothy Hodgkin Fellowship funded by the Royal Society, and acknowledges start-up support from the University of Manchester and the University of Sheffield.
The Theoretical Astrophysics Program (TAP) at the University of Arizona provided resources that supported this work.

\end{acknowledgments}

\appendix

\section{Solar focusing and wake-template construction}
\label{app:wake_focusing}

This appendix collects the focusing relations~
\cite{Sikivie:2002bj,Alenazi:2006wu,Lee:2013wza,Kim:2021yyo} used to construct the physical profiles summarized in Sec.~\ref{sec:wake}. The equations below define the
particle and wave templates; mission geometry and the differential-ranging response are treated in the main text.

\subsection{Asymptotic component and coordinate convention}
\label{app:wake_coordinates}

Let $f_\infty(\mathbf u)$ be the asymptotic velocity distribution of one
unbound dark-matter component in the solar rest frame, normalized by
\begin{equation}
    \int \mathrm d^3u\,f_\infty(\mathbf u)=1.
    \label{eq:velocity_distribution_normalization}
\end{equation}
Its mean velocity is
\begin{equation}
    \mathbf v_\infty
    \equiv
    \int \mathrm d^3u\,\mathbf u f_\infty(\mathbf u)
    =
    v_\infty\hat{\mathbf w}.
    \label{eq:asymptotic_velocity}
\end{equation}
Thus $\hat{\mathbf w}$ points downstream, in the direction of propagation of
the selected component. A numerical value of $(\theta_w,\phi_w)$ must be
obtained by transforming a specified solar-frame flow vector into a specified
ecliptic frame. The inclination between the Galactic and ecliptic planes does
not determine the wake direction by itself.

The focused density of the selected component is
\begin{equation}
    \rho_{\rm comp}^{\odot}(\mathbf r)
    =
    \rho_{\rm comp}\left[1+\delta_w(\mathbf r)\right],
    \label{eq:component_focused_density}
\end{equation}
so that its contribution to the gravitational wake template is
$\delta\rho_w=\rho_{\rm comp}\delta_w$. If several asymptotic components are
included, their excess densities add linearly.

\subsection{Particle focusing}
\label{app:particle_focusing}

A hyperbolic trajectory may be labelled by its asymptotic speed $u$ and
impact parameter $b_\odot$. In the point-mass approximation, its deflection is
\begin{equation}
    \alpha_\odot(b_\odot,u)
    =
    2\arctan\left(
        \frac{G\Msun}{b_\odot u^2}
    \right)
    \simeq
    \frac{2G\Msun}{b_\odot u^2},
    \label{eq:deflection}
\end{equation}
where the final form assumes weak deflection. The associated axial crossing
satisfies
\begin{equation}
    s_f
    \simeq
    \frac{b_\odot}{\alpha_\odot}
    \simeq
    \frac{b_\odot^2u^2}{2G\Msun},
    \qquad
    b_\odot(s_f)
    \simeq
    \left(
        \frac{2G\Msun s_f}{u^2}
    \right)^{1/2}.
    \label{eq:focus_distance}
\end{equation}
This relation is purely geometric: it identifies where a ray bundle converges,
but it does not determine the density carried by that bundle.

The density follows from collisionless phase-space transport. Let
$\mathbf u_\infty(\mathbf r,\mathbf v)$ denote the asymptotic velocity found
by integrating the local phase-space point $(\mathbf r,\mathbf v)$ backward
along its unbound solar orbit. Conservation of phase-space density then gives
\begin{equation}
    1+\delta_w^{\rm part}(\mathbf r)
    =
    \int \mathrm d^3v\,
    f_\infty\!\left[
        \mathbf u_\infty(\mathbf r,\mathbf v)
    \right].
    \label{eq:particle_focusing_integral}
\end{equation}
This backward map defines the particle template used in a physical forecast
\cite{Sikivie:2002bj,Alenazi:2006wu,Lee:2013wza}. The single-ray focal relation
alone is insufficient. Orbits that enter the Sun must use the enclosed solar
mass profile instead of a point source.

For quick checks of a numerically constructed Maxwellian template, we use two
scalar summaries. With one-dimensional dispersion $\sigma$ and bulk speed
$v_\infty$, the downstream-axis contrast is approximately~\cite{Kim:2021yyo}
\begin{equation}
    1+\delta_{\rm ds}
    \simeq
    \left(
        1+\frac{v_e^2(r)}{\sigma^2}
    \right)^{1/2},
    ~~~
    v_e^2(r)=\frac{2G\Msun}{r},
    \label{eq:downstream_contrast}
\end{equation}
while the solid-angle average is approximately
\begin{equation}
    1+\delta_{\rm avg}
    \simeq
    \left(
        1+\frac{v_e^2(r)}{\sigma^2+v_\infty^2}
    \right)^{1/2}.
    \label{eq:avg_contrast}
\end{equation}
These relations are diagnostics for scale and normalization; the trajectory
calculation uses the full three-dimensional phase-space map.

\subsection{Wave focusing}
\label{app:wave_focusing}

Each incident wave mode is propagated at its own asymptotic speed $u$.
Energy conservation in the solar potential gives the local speed
\begin{equation}
    \widetilde u^2(r)=u^2+v_e^2(r).
    \label{eq:focused_mode_speed}
\end{equation}
The corresponding de Broglie wavelength and reduced wavelength are
\begin{equation}
\begin{aligned}
    \lambda_{\rm dB}(r;u)
    &=
    \frac{2\pi\hbar}{m\widetilde u(r)}
    \\
    &\simeq
    10.4\,\AU
    \left(
        \frac{10^{-15}\,\mathrm{eV}}{m}
    \right)
    \left(
        \frac{240\,\mathrm{km\,s^{-1}}}{\widetilde u(r)}
    \right),
\end{aligned}
    \label{eq:debroglie}
\end{equation}
\begin{equation}
\begin{aligned}
    \ell_{\rm dB}(r;u)
    &\equiv
    \frac{\lambda_{\rm dB}}{2\pi}
    =
    \frac{\hbar}{m\widetilde u(r)}
    \\
    &\simeq
    1.65\,\AU
    \left(
        \frac{10^{-15}\,\mathrm{eV}}{m}
    \right)
    \left(
        \frac{240\,\mathrm{km\,s^{-1}}}{\widetilde u(r)}
    \right).
\end{aligned}
    \label{eq:reduced_debroglie}
\end{equation}
A component with a velocity distribution contains many such scales; a single
$\widetilde u$ is used below only for numerical orientation, whereas the
focused template is evaluated mode by mode.

A convenient measure of the wave regime is the number of reduced de Broglie
lengths across the sampled radius,
\begin{equation}
    x(r;u)
    \equiv
    \frac{m\widetilde u(r)r}{\hbar}
    =
    \frac{r}{\ell_{\rm dB}(r;u)}.
    \label{eq:x_wave_parameter}
\end{equation}
For $x\gg1$, many phase oscillations fit inside the scale $r$ and a
coarse-grained density approaches the particle result. For $x\lesssim1$,
diffraction remains resolved and the focused profile may contain both positive
and negative contrast \cite{Kim:2021yyo}. Setting $x=1$ at a chosen radius
defines the crossover mass
\begin{equation}
\begin{aligned}
    m_{\rm wo}(r_\ast)
    &\equiv
    \frac{\hbar}{\widetilde u(r_\ast)r_\ast}
    \\
    &\simeq
    1.6\times10^{-15}\,\mathrm{eV}
    \left(
        \frac{240\,\mathrm{km\,s^{-1}}}{\widetilde u(r_\ast)}
    \right)
    \left(
        \frac{1\,\AU}{r_\ast}
    \right).
\end{aligned}
    \label{eq:m_res}
\end{equation}
Thus a survey over $1$--$10\,\AU$ samples characteristic crossover masses of
order
\begin{equation}
    m\sim10^{-16}\text{--}10^{-15}\,\mathrm{eV}
    \label{eq:wave_mass_range}
\end{equation}
for the reference speed. This mass range identifies the physical wake regime
sampled by the stated radius and reference speed.

The ensemble-averaged wave contrast is obtained by weighting the propagated
mode intensities with the asymptotic velocity distribution,
\begin{equation}
    1+\delta_w^{\rm wave}(\mathbf r)
    =
    \int \mathrm d^3u\,
    f_\infty(\mathbf u)
    \left|\psi_{\mathbf u}(\mathbf r)\right|^2.
    \label{eq:wave_contrast_direct}
\end{equation}
This construction must be performed independently of the particle map; in
general, no position-independent factor converts one template into the other
\cite{Kim:2021yyo}.

Near the downstream axis, two broadening mechanisms compete: the directional
spread of the incident component and wave diffraction. For a narrow flow,
their angular support can be summarized as
\begin{equation}
    \Delta\mu
    \sim
    \max\left[
        \frac{\ell_{\rm dB}}{s},
        \left(\frac{\sigma}{v_\infty}\right)^2
    \right],
    \qquad
    \mu\equiv\hat{\mathbf r}\cdot\hat{\mathbf w},
    \label{eq:wave_angular_support}
\end{equation}
following the scaling in Ref.~\cite{Kim:2021yyo}. Converting this angular
support into transverse distance gives the order-of-magnitude scales
\begin{equation}
\begin{aligned}
    w_\sigma(s)
    &\sim
    s\frac{\sigma}{v_\infty},
    \\
    w_{\rm dif}(s)
    &\sim
    \min\left[
        s,\sqrt{s\ell_{\rm dB}}
    \right],
    \\
    w_{\rm phys}(s)
    &\sim
    \max\left[
        w_\sigma(s),w_{\rm dif}(s)
    \right].
\end{aligned}
    \label{eq:wake_width}
\end{equation}
These expressions guide targeting estimates only. The width used in a
forecast must be measured from the computed density or projected-Hessian map
with an explicit convention, such as a half-maximum radius or an
integrated-contrast interval.

Pointing and navigation uncertainty do not broaden the physical wake. If
$\sigma_{\rm nav}$ is the transverse navigation uncertainty and
$\sigma_\theta$ is the uncertainty in the wake direction, their contribution
to the targeting uncertainty may be written as
\begin{equation}
    \sigma_b^2(s)
    =
    \sigma_{\rm nav}^2+s^2\sigma_\theta^2,
    \qquad
    w_{\rm targ}^2(s)
    =
    w_{\rm phys}^2(s)+\sigma_b^2(s).
    \label{eq:targeting_width}
\end{equation}
Only $w_{\rm phys}$ enters the density and gravity calculation;
$w_{\rm targ}$ is used for acquisition and trajectory robustness.

\subsection{Benchmark and implementation conventions}
\label{app:wake_benchmarks}

We use
\begin{equation}
\begin{aligned}
    \rho_{\rm ref}
    &\equiv
    0.4\,\mathrm{GeV\,cm^{-3}}
    \\
    &=
    7.13\times10^{-22}\,\mathrm{kg\,m^{-3}}
    \\
    &=
    1.20\times10^{-18}\,\Msun\,\AU^{-3}
\end{aligned}
    \label{eq:rho_local_units}
\end{equation}
as a reference local-density normalization. The wake excess is specified
separately. If a component carrying this full reference density has a one-percent
peak contrast, then
\begin{equation}
    \delta\rho_{\rm pk}
    =
    4.0\times10^{-3}\,\mathrm{GeV\,cm^{-3}}
    =
    1.20\times10^{-20}\,\Msun\,\AU^{-3}.
    \label{eq:halo_overdensity_number}
\end{equation}
For a stream or dark-disk benchmark, $\rho_{\rm comp}$ and the focused
contrast must come from the same phase-space model. They should not be
maximized independently.

For analytic checks and software validation only, a finite Gaussian excess may
be written as
\begin{equation}
    \delta\rho_w^{\rm toy}(s,\mathbf r_\perp)
    =
    \delta\rho_{\rm pk}
    \exp\left[-\frac{r_\perp^2}{2w^2}\right]
    \exp\left[-\frac{(s-s_0)^2}{2\ell^2}\right].
    \label{eq:resonant_wake_profile}
\end{equation}
This profile is useful for checking linear response and finite-baseline
corrections, and it also exposes parameter degeneracies.
Physical reach must be calculated with the particle or wave profiles above.

The potential of any finite excess profile is
\begin{equation}
    \Phi_w(\mathbf r)
    =
    -G\int \mathrm d^3r'\,
    \frac{\delta\rho_w(\mathbf r')}
    {|\mathbf r-\mathbf r'|},
    \label{eq:wake_potential_integral}
\end{equation}
with
\begin{equation}
    \mathcal E^w_{ij}
    =
    \partial_i\partial_j\Phi_w,
    \qquad
    \mathcal E^w_{ii}
    =
    4\pi G\,\delta\rho_w.
    \label{eq:wake_hessian_trace}
\end{equation}
Poisson's equation fixes only the trace of the Hessian. The component measured
by a spacecraft pair depends on the full three-dimensional profile and the
instantaneous baseline orientation.

\bibliographystyle{apsrev4-2}
\bibliography{references}

\end{document}